\documentclass[12pt]{article}
\usepackage{epsfig,amsfonts,amsthm}
\usepackage{amsmath,amssymb}
\newcommand{\be}{\begin{equation}}
\newcommand{\ee}{\end{equation}}
\newcommand{\bea}{\begin{eqnarray}}
\newcommand{\eea}{\end{eqnarray}}

\newcommand{\fd}{\phi^\dagger}
\renewcommand{\Re}{\mathrm{Re }}
\renewcommand{\Im}{\mathrm{Im }}

\newcommand{\doublet}[2]{ \left( \begin{array}{c}#1 \\ #2 \end{array}\right) }

\newcommand{\lr}[1]{ \langle #1 \rangle}
 \newcommand{\Tr}{\mathrm{Tr}}

\providecommand{\RR}{{\mathbb{R}}}

\def\lsim{\mathrel{\rlap{\lower4pt\hbox{\hskip1pt$\sim$}}
    \raise1pt\hbox{$<$}}}         
\def\gsim{\mathrel{\rlap{\lower4pt\hbox{\hskip1pt$\sim$}}
    \raise1pt\hbox{$>$}}}         

\title{Properties of the general NHDM. II. Higgs potential and its symmetries}

\author{I.~P.~Ivanov
\\
  {\small IFPA, Universit\'{e} de Li\`{e}ge, All\'{e}e du 6 Ao\^{u}t 17, b\^{a}timent B5a, 4000 Li\`{e}ge, Belgium}\\
  {\small and}\\
  {\small  Sobolev Institute of Mathematics, Koptyug avenue 4, 630090, Novosibirsk, Russia}\\
  }

\begin{document}
\maketitle

\begin{abstract}
We continue our analysis of the general $N$-Higgs-doublet model and focus
of the Higgs potential description in the space of gauge orbits. 
We develop a geometric technique that allows us 
to study the global minimum of the potential without explicitly finding its position.
We discuss symmetry patterns of the NHDM potential, and illustrate
the general discussion with various specific variants of the
three-Higgs-doublet model. 
\end{abstract}

\section{Introduction}

The electroweak theory relies on the Higgs mechanism of the electroweak symmetry breaking (EWSB). 
Many variants of the Higgs mechanism have already been suggested, see e.g. \cite{CPNSh}.
They range from conceptually simple ones where one just adds extra particles 
to the scalar sector of the Standard model to the models involving new gauge interactions, 
extra dimensions at the TeV scale, supersymmetry, etc. 
It is not presently known what particular variant is realized in Nature,
and it is the key goal of the LHC and of future linear colliders to answer that question.
However, in order to safely interpret possible manifestations 
of the physics beyond the Standard model (BSM) in the future experimental data,
it is necessary to clearly understand in advance what each model is capable of.

There is a broad class of BSM models known as $N$-Higgs-doublet models (NHDM),
in which one gets New Physics just by introducing several copies of a Higgs doublet
with the usual electroweak quantum numbers.
They are among the most conservative extensions of the Standard model and are driven
by the idea to take the notion of ``generations'' to the scalar sector.
This idea is quite natural because, first, there is no strong argument that would
favor a single Higgs doublet, second, we do have generations of fermions without understanding their origin,
and third, since fermion masses and mixing patterns result from their Yukawa interactions
with scalars, there might exist a link between generations of scalars and fermions.
Additional support is lent by supersymmetric models and certain low-energy 
realizations of superstring/brane models, which require presence of several Higgs generations,
see \cite{gupta2010} and references therein.

Multi-Higgs-doublet models were studied a lot in the past decades (see some references below). 
Virtually all the analyses were restricted to simple variants of the Higgs self-interaction potential, 
with interesting phenomenological consequences observed. However one cannot know if these particular
cases exhaust all the New Physics offered by several doublets or if there are some ``hidden treasures''
lurking in the multi-dimensional parameter space of the model.
To find them, one certainly needs to study the $N$-doublet models with most general 
scalar potential and Yukawa sector.

Algebraically, working with the most general $N$-Higgs-doublet potential is a hard task.
Difficulties arise even at the tree level within the scalar sector of the theory.
A very representative case is given by the two-Higgs-doublet model (2HDM), \cite{CPNSh,TDLee,Hunter}.
Here, the Higgs potential cannot be minimized explicitly in the general case.
As a result, for a long time only relatively simple variants of the 2HDM remained analyzed,
while the most general 2HDM stood barely touched. 
However, in the last several years new tools have been developed
which led to many insights into the properties of the general 2HDM.
These methods are based on the idea of the {\em reparametrization invariance}
of the physical observables in the scalar sector: 
a unitary transformation between the Higgs doublets changes the parameters 
of the lagrangian, but the physical observables such as the Higgs boson masses 
and the sum of squares of the vacuum expectation values remain the same.
This idea was implemented via the tensorial formalism at the level of Higgs fields \cite{CP,haber,Branco:2005em,haber2,oneil} 
or via geometric constructions in the space of gauge-invariant bilinears 
\cite{sartori,nagel,heidelberg,nishi2006,ivanov0}.
In the latter case the formalism was extended to include non-unitary reparametrization 
transformations \cite{ivanov1,ivanov2,nishi2008},
which revealed interesting geometric properties of the 2HDM in the orbit space equipped
with the Minkowskian metric.

This formalism led to discovery of new features of the scalar sector of 2HDM 
that could not be found or were just too obscure to be observed with more tranditional techniques.
They include, among others, the following results:
\begin{itemize}
\item There can exist at most two distinct minima of the Higgs potential, \cite{ivanov2};
\item charge-breaking and neutral minima never coexist (although this fact 
was known since \cite{ferreira2004CB}, its proof becomes straightforward in the Minkowskian formalism, \cite{ivanov1});
\item $CP$-conserving and $CP$-violating minima cannot coexist, \cite{ivanov1};
\item there are exactly six conjugacy classes of symmetries that can be imposed on the scalar lagrangian \cite{heidelberg,nishi2006,ivanov0},
all specific realizations of each symmetry can be mapped onto each other by a reparametrization transformation.
Remarkably, this conclusion holds even for the scalar potential alone, despite the fact that it can be more symmetric than the 
full scalar lagrangian, \cite{ivanov2}; 
\item the maximal breaking of a discrete symmetry of the potential consists in removing just a single $Z_2$ factor, \cite{ivanov2};
\item the full tree-level phase diagram of the scalar sector was described and possible phase transitions were identified \cite{ivanov2};
\item the thermal evolution of this phase diagram was calculated in the first non-trivial approximation, 
with interesting results for possible multiple phase transition paths of the early Universe, \cite{thermal}.
\end{itemize}
Apart from that, many algebraic conditions were derived in a basis-invariant form (positivity constraints,
conditions for existence of any specific symmetry in a given lagrangian, 
conditions when this symmetry is broken).

It is a natural idea to extend these successful techniques to $N$ doublets.
The general NHDM is obviously more involved than 2HDM, both at the level of
scalar sector and Yukawa interactions, see examples in \cite{weinberg1976,Branco1980,DeshpandeHe1994}. 
Some properties of the general NHDM were analyzed in 
\cite{heidelberg,Erdem1995,barroso2006,nishi:nhdm,ferreira-nhdm,zarrinkamar},
with a special emphasis on $CP$-violation, \cite{nishi2006,Branco:2005em,lavoura1994}.
However, a method to systematically explore all the possibilities offered with $N$ doublets
was still missing.

It is the aim of this paper to fill this gap, at least partially.
Making use of the results of the companion paper \cite{NHDM2010}, 
we study here the general NHDM Higgs potential in the space of bilinears.
We develop a geometric technique that allows us to study the properties 
of the minima of the Higgs potential without explicitly finding its position.
The general constructions are illustrated with several
variants of the three-Higgs-doublet model (3HDM) displaying various patterns 
of spontaneous symmetry breaking.
Note that in this paper we focus only on the Higgs potential; the interaction between 
the Higgs sector and the fermions is an intriguing topic, which we do not touch here.

The paper is organized as follows. In Section 2 we briefly overview some of the results 
obtained in \cite{NHDM2010} concerning the orbit space of the NHDM.
Then, in Section 3 we describe the equipotential surface technique.
Section 4 is devoted to symmetries of NHDM. These general discussions are used 
in Section 5 to construct several variants of 3HDM with various symmetries and symmetry breaking
patterns. We conclude the paper with a discussion of the results and an outline for future work.

\section{The orbit space of NHDM}\label{section:orbit}

\subsection{General $N$}

A detailed description of the orbit space of NHDM was given in \cite{NHDM2010}.
Here we just introduce the notation and briefly summarize some of its results.

In the $N$-Higgs-doublet model we introduce $N$ Higgs doublets $\phi_a$, $a=1,\dots,N$.
The general renormalizable Higgs potential of NHDM is constructed from the gauge-invariant
combinations $(\phi_a^\dagger \phi_b)$, which describe the gauge orbits in the Higgs space. 
Following \cite{nagel,heidelberg,nishi2006,nishi:nhdm}, we organize them in the $K$-matrix:
$K_{ab} = (\phi^\dagger_b\phi_a)$,
which is a hermitian and positive-semidefinite $N\times N$ matrix with rank $\le 2$.
A rank-2 $K$-matrix corresponds to a charge-breaking minimum, while a rank-1 matrix corresponds
to a neutral minimum. Any $K$-matrix can be brought to a diagonal form by an appropriate 
reparametrization transformation (a unitary or antiunitary rotation among the doublets).

The $K$-matrix can be decomposed as
\be
\label{K:decomposed}
K \equiv r_0 \cdot \sqrt{{2 \over N(N-1)}}{\bf 1}_N + r_i \lambda_i\,, \quad 
r_0 = \sqrt{{N-1 \over 2N}}\Tr K\,,\quad r_i = {1\over 2}\Tr[K\lambda_i]\,,
\quad i = 1, \dots , N^2-1\,.
\ee
Here ${\bf 1}_N$ is the $N\times N$ unit matrix and $\lambda_i$ are generators of the algebra $su(N)$.
The coefficient in front of the unit matrix in (\ref{K:decomposed}) is chosen for future convenience.
Eq.~(\ref{K:decomposed}) defines a linear and invertible map from the space of $K$-matrices to the space of 
real vectors\footnote{The notation $r^\mu$ alludes to the Minkowski-space formalism of 
\cite{ivanov1,ivanov2,nishi2006,nishi2008}.
Since we limit ourselves only to the (anti)unitary reparametrization transformations,
we will not use this formalism in this paper;
$r^\mu$ is just a short notation for $(r_0,r_i)$.} 
$r^\mu \equiv (r_0, r_i) \in \RR^{N^2}$, which we call the {\em adjoint space}. 
As it was noted in \cite{heidelberg,nishi:nhdm}, this map is not surjective.
Therefore, the orbit space does not cover the entire space of vectors $r^\mu$, 
but is represented by a certain manifold embedded in it, which we denote by ${\cal V}_\Phi$.
This manifold can be described algebraically with a list of polynomial (in)equalities in components
of $r^\mu$.
This list starts with 
\be
\label{sun:constraints}
r_0 \ge 0\,, \qquad r_0^2 - \vec r^2 \ge 0\,,\qquad 
d_{ijk}r_ir_jr_k + {N-2 \over\sqrt{2N(N-1)}}(r_0^2-3\vec r^2) = 0\,,
\ee 
where $d_{ijk}$ are the fully symmetric $SU(N)$ tensors,
see details in \cite{NHDM2010}. 
The manifold of the neutral vacua can be obtained from (\ref{sun:constraints}) 
by turning the second inequality into equality.
In 2HDM we have just the first two inequalities, which
define the surface and interior of the forward lightcone in $\RR^4$. In 3HDM the third equality is added, 
which carves out a specific 8-dimensional manifold in the forward lightcone in $\RR^9$.

Some geometric properties of the orbit space manifold ${\cal V}_\Phi$ were established in \cite{NHDM2010}.
The dimensionalities of the charge-breaking and neutral parts of the orbit space manifold, 
${\cal V}_C$ and ${\cal V}_N$, are $4N-4$ and $2N-1$, respectively. 
Since all the equations in (\ref{sun:constraints}) are homogeneous, 
the orbit space has a conical shape in the adjoint space. Its base (i.e. a $r_0=\mathrm{const}$ section) can be
described in the $(N^2-1)$-dimensional vector space of $\vec n \equiv \vec r / r_0$, 
where ${\cal V}_N$ becomes the complex projective space $\mathbb{CP}^{N-1}$ embedded in $\RR^{N^2-1}$,
while the entire orbit space ${\cal V}_\Phi$ can be reconstructed as the symmetric join of ${\cal V}_N$
(the union of all points that lie on line segments drawn between all possible pairs of ${\cal V}_N$ points).

It is also very useful to consider the {\em root space} of NHDM.
It is the $(N-1)$-dimensional subspace spanned by the Cartan subalgebra of $su(N)$,
i.e. the space of coordinates $n_i$ in front of the diagonal matrices $\lambda_i$ in (\ref{K:decomposed}),
with all the other coordinates set to zero.
Any point of the orbit space can be brought to the root space by an appropriate reparametrization transformation.
In the root space, the orbit space is represented by an $(N-1)$-simplex:
the neutral orbit space corresponds to the $N$ vertices of an $N$-dimensional regular tetrahedron,
while the charge-breaking orbit space corresponds to the $N(N-1)/2$ edges joining all pairs of vertices.
Note that the interior of the tetrahedron does not belong to the orbit space, so that
a curious feature of the NHDM is the existence of a ``hole'' inside the orbit space (or an empty ``inner cone''
in the $r^\mu$ space).

\subsection{The case of 3HDM}

Let us consider more closely the orbit space of the three-Higgs-doublet model (3HDM).
For $N=3$, the matrices $\lambda_i$, $i=1,\dots,8$, are the standard Gell-Mann matrices,
and the Cartan subalgebra is formed by the diagonal matrices  
\be
\lambda_3 = \mathrm{diag}(1,-1,0)\quad \mbox{and} \quad\lambda_8 = {1 \over \sqrt{3}}\mathrm{diag}(1,1,-2).
\ee
The explicit expressions for the field bilinears are:
\bea
&& r_0 = {1 \over\sqrt{3}} \Tr K = {(\phi_1^\dagger\phi_1) + (\phi_2^\dagger\phi_2) + (\phi_3^\dagger\phi_3)\over\sqrt{3}}\,.\\
&& 
r_i = {1\over 2}\Tr[K \lambda_i]\,, \quad r_3 = {(\phi_1^\dagger\phi_1) - (\phi_2^\dagger\phi_2) \over 2}\,,\quad
r_8 = {(\phi_1^\dagger\phi_1) + (\phi_2^\dagger\phi_2) - 2(\phi_3^\dagger\phi_3) \over 2\sqrt{3}}\,,\quad
\nonumber\\
&&r_1 = \Re(\phi_1^\dagger\phi_2)\,,\ r_2 = \Im(\phi_1^\dagger\phi_2)\,,\
r_4 = \Re(\phi_1^\dagger\phi_3)\,,\nonumber\\[2mm] &&r_5 = \Im(\phi_1^\dagger\phi_3)\,,\
r_6 = \Re(\phi_2^\dagger\phi_3)\,,\ r_7 = \Im(\phi_2^\dagger\phi_3)\,.\nonumber
\eea
It is also useful to group the last six real coordinates into three ``complex coordinates'':
\be
r_{12} = r_1 + i r_2\,,\quad r_{45} = r_4 - i r_5\,,\quad r_{67} = r_6 + i r_7\,,\label{complexcoordinates}
\ee 
with the same labels for the normalized vectors $n_i \equiv r_i/r_0$.

\begin{figure}[!htb]
   \centering
\includegraphics[height=6cm]{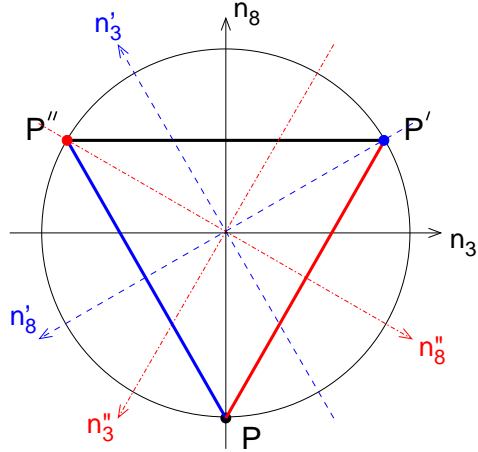}
\caption{The root plane for 3HDM. Shown are the unit circle (section of the lightcone),
the three points $P$, $P'$, $P''$ representing the neutral manifold, the three line segments
representing the charge-breaking manifold, as well as the three sets of 
coordinates: $(n_3,n_8)$ shown in black solid lines, 
$(n'_3,n'_8)$ shown in dashed blue lines, and $(n''_3,n''_8)$ shown in dash-dotted red lines.}
   \label{fig-n3n8plane}
\end{figure}

The root space of 3HDM is given by the $(n_3,n_8)$ plane, shown in Fig.~\ref{fig-n3n8plane}.
On the root plane, the orbit space is represented by three points $P$, $P'$, $P''$ (neutral manifold)
and the three line segments joining them (the charge-breaking manifold). 

The full orbit space ${\cal V}_\Phi$ is symmetric under the group $S_3$ of permutations of the three doublets, but the description in terms
of $n_i$ breaks this symmetry. In order to restore it, we introduce two extra sets of the coordinates,
$n'_i$ and $n''_i$, which are obtained from $n_i$ by cyclic permutations of the three doublets 
$\phi_1 \to \phi_2 \to \phi_3\to\phi_1$ and its inverse:
\bea
&&
n'_3 = -{1\over 2} n_3 + {\sqrt{3} \over 2} n_8\,,\quad n'_8 = - {\sqrt{3}\over 2} n_3 - {1 \over 2} n_8\,,\quad
n'_{12} = n_{67}\,,\quad n'_{45} = n_{12}\,,\quad n'_{67} = n_{45}\,,\nonumber\\
&&
n''_3 = -{1\over 2} n_3 - {\sqrt{3} \over 2} n_8\,,\quad n''_8 = {\sqrt{3}\over 2} n_3 - {1 \over 2} n_8\,,\quad
n''_{12} = n_{45}\,,\quad n''_{45} = n_{67}\,,\quad n''_{67} = n_{12}\,. \label{three-sets}
\eea
Note that $n_3+n'_3+n''_3=n_8+n'_8+n''_8=0$.
In terms of these coordinates, the three neutral points are
\be
P:\ n_3 = 0,\, n_8 = -1\,,\quad
P':\ n'_3 = 0,\, n'_8 = -1\,,\quad
P'':\ n''_3 = 0,\, n''_8 = -1\,.\label{pointsPPP2}
\ee
Finally, we also introduce the baricentric coordinates
\be
p ={1-2n_8 \over 3}\,,\quad p'={1-2n_8' \over 3}\,,\quad p''={1-2n_8'' \over 3}\,, \quad\mbox{with}\quad p+p'+p''=1\,,
\label{baricentric}
\ee
which are proportional to the distances from a given point to
the three edges of the triangle. 
Using all these coordinates, one can write the $K$-matrix of 3HDM in a very symmetric way:
\be
K_{ab} = r_0 \left(
\begin{array}{ccc} 
\sqrt{3}p' &  n_{12}^* & n_{12}'' \\[2mm]
n_{12} & \sqrt{3}p'' & n_{12}^{\prime *} \\[2mm]
n_{12}^{\prime\prime *} &  n_{12}' & \sqrt{3}p
\end{array}
\right)\,.\label{Kmatrixsymm}
\ee

Let us now turn to the third algebraic conditions in (\ref{sun:constraints}), which we will refer to as ``the $d$-condition'':
\be
\label{dcondition}
\sqrt{3}d_{ijk} n_i n_j n_k = {3 \vec n^2 - 1\over 2}\,.
\ee
With the three sets of coordinates, this condition can be rewritten symmetrically as
\be
\label{d:bari}
p|n_{12}|^2 + p'|n'_{12}|^2 + p'' |n''_{12}|^2 - 3 p p' p'' - {2 \over \sqrt{3}}\Re(n_{12}n'_{12}n''_{12}) = 0\,.
\ee
For the neutral manifold, the $d$-condition can be simplified further. In this case we have 
\be
3 p' p'' = |n_{12}|^2\,,\quad  3 p p' = |n''_{12}|^2\,,\quad 3 p'' p = |n'_{12}|^2 \,.\label{zab:3HDM}
\ee
Denoting the sum of the phases of $n_{12}$, $n'_{12}$, and $n''_{12}$ by $\gamma$,
one can cast the $d$-condition for the neutral orbit space into
\be
p p' p''(1-\cos\gamma) = 0\,.\label{shape5}
\ee

\section{Minimization of the general NHDM Higgs potential}\label{section:potential-NHDM}

The key attractive feature of the adjoint space formalism is the simplicity with which
one can describe the Higgs potential and discuss its symmetries.
The most general gauge-invariant and renormalizable Higgs potential
with $N$ Higgs doublets can be written in a very compact way
\be
V = - M_\mu r^\mu + {1 \over 2}\Lambda_{\mu\nu} r^\mu r^\nu\,.\label{potential}
\ee
As we mentioned above, we do not use the Minkowski-space formalism of \cite{ivanov1,ivanov2,nishi2006,nishi2008}
in this paper, so $r^\mu$ just stands for $(r_0,r_i)$, $i = 1,\dots, N^2-1$, and all index contractions are
meant to be Euclidean:
\be
\label{potential:again}
V = - M_0 r_0 - M_i r_i + {1 \over 2}\Lambda_{00} r_0^2 + \Lambda_{0i} r_0 r_i + {1 \over 2}\Lambda_{ij} r_i r_j\,.
\ee
Note that when constructing a potential, we have the full freedom to choose $N^2$ components of $M_\mu$ 
and $N^2(N^2+1)/2$ components of $\Lambda_{\mu\nu}$.

The first thing one needs to do with the potential is to minimize it.
Unfortunately, this task cannot be done with straightforward algebra for a general potential.
Alternatively, one can start from a minimum and construct a potential, which by construction has a minimum at a chosen point.
In this case, however, one must be sure that the potential does not possess another, deeper minimum, 
which again cannot be verified with straightforward algebra.

In \cite{ivanov2} a geometric technique was suggested for 2HDM that allowed one to study the properties of the global minima
of the Higgs potential without knowing its exact position.
With this technique, one constructs equipotential surfaces and finds their intersection with the orbit space.
Although this approach did not help to explicitly minimize a given potential,
it offered a good understanding of the entire spectrum of possibilities (symmetries and their breaking,
coexistence of minima, phase transitions), which can in principle arise at the tree-level in 2HDM.  

In this Section we generalize this approach to $N$ doublets.

\subsection{Equipotential surfaces}

The Higgs potential (\ref{potential}) is a quadratic form in the components of $r^\mu$.
In order to be physically realizable in terms of doublets, $r^\mu$ must lie in the orbit space,
whose complicated shape we briefly described in Section~\ref{section:orbit}.
Let us, however, put aside this requirement for a moment and consider 
the quadratic form (\ref{potential:again}) in the entire adjoint space $\RR^{N^2}$.

We choose a real number $C$ and construct an {\em equipotential surface} ${\cal M}_C$ as the set of 
all points $r^\mu$ satisfying $V(r^\mu) = C$. Taking another value of $C$ will result in another 
equipotential surface constructed from the same potential, and so on. 
Several observations can be inferred from this definition.
\begin{itemize}
\item
Each ${\cal M}_C$ is an $(N^2-1)$-quadric, i.e. a second-order algebraic manifold embedded in $R^{N^2}$: a hyperboloid, an ellipsoid, etc.
If the potential has flat directions, then the equipotential surfaces are cylindrical along these directions. 
\item
Two equipotential surfaces of the same potential but with different $C_i$ do not intersect.
\item
The equipotential surfaces are linearly ordered: if $C_1<C_2<C_3$, then ${\cal M}_{C_2}$ lies between 
${\cal M}_{C_1}$ and ${\cal M}_{C_3}$. One can also say that ${\cal M}_{C_2}$ lies {\em higher} than ${\cal M}_{C_1}$,
but {\em lower} than ${\cal M}_{C_3}$. 
\end{itemize}
Thus, a given Higgs potential can be associated with a nested {\em family of equipotential surfaces}, 
which covers the entire adjoint space.
There are two sorts of families of equipotential surfaces to be considered:
\begin{itemize}
\item
$R$-family: solutions to equation $V(r)=C$ exist for any real $C$.
\item
$R^+$-family: solutions to equation $V(r)=C$ exist only for all real $C$ larger or equal to some value.  
\end{itemize}
Which case we have depends solely on the signs of the Euclidean eigenvalues of $\Lambda_{\mu\nu}$. 
The quadratic form
\be
\label{quadraticform}
\Lambda_{\mu\nu}r^\mu r^\nu \equiv \Lambda_{00} r_0^2 + 2 \Lambda_{0i}r_0 r_i + \Lambda_{ij}r_i r_j
\ee
is real and symmetric, therefore, it can be diagonalized in the $(r_0,\vec r)$ space by some rotation
of the coordinates: $(r_0,\vec r) \to \{r_q\}$, $q = 1,\dots, N^2$. 
After this transformation, the quadratic form (\ref{quadraticform}) becomes $\Lambda_q r_q^2$, 
with at least one positive eigenvalue among $\Lambda_q$ (otherwise, the potential would not be bounded from below
in the orbit space).
The $R^+$-family corresponds to the case when all $\Lambda_q \ge 0$,
that is when $\Lambda_{\mu\nu}$ is positive semidefinite in the entire adjoint space $\RR^{N^2}$.
The $R$-family corresponds to situations when at least one of $\Lambda_q$ is negative.

Note that we do not claim that a rotation diagonalizing $\Lambda_{\mu\nu}$ is physical and can be realized by a certain
transformation of doublets. It well may be not because there is no transformation of the doublets that mixes $r_0$ and $r_i$
keeping $r_0^2+r_i^2$ constant. We use this unphysical transformation just to characterize the 
spectrum of the matrix $\Lambda_{\mu\nu}$ and define the shape of the equipotential surfaces.

\subsection{Positivity conditions}

The physical Higgs potentials we consider must be stable, i.e. bounded from below.
Potentials with a quadratic $V_2$ and a quartic $V_4$ terms can be stable in two senses.
Using terminology of \cite{heidelberg}, we say that a potential is stable in a strong sense 
if its $V_4$ increases along any ray in the space of Higgs fields.
A potential stable in a weak sense is the one whose $V_4$ can have flat directions, 
but $V_2$ grows along them.
In this paper we focus on potentials stable in a strong sense. Potentials stable in a weak sense
can be analyzed in a similar manner.

An $R^+$-family of equipotential surfaces automatically guarantees the stability of the potential.
For an $R$-family, we require that the quartic part of the potential be positive definite
everywhere in the orbit space (except the origin):
\be
\Lambda_{\mu\nu} r^\mu r^\nu > 0 \quad \mbox{for all non-zero $r^\mu$ in the orbit space}\,.
\ee
Unfortunately, we could not find a compact algebraic reformulation of this criterion in
terms of components of $\Lambda_{\mu\nu}$ only (an attempts to find such conditions
using hyperspherical coordinates can be found in \cite{zarrinkamar}). 
It is possible that such a criterion
will involve its Minkowskian eigenvalues, similarly to the result \cite{ivanov1,nishi2008} for 2HDM,
although a sufficient condition is easy to formulate for NHDM as well, see \cite{nishi:nhdm}.

On the other hand, the positivity constraint on the Higgs potential can be easily described geometrically,
which helps visualize the freedom we have when constructing NHDM potentials.
Consider again the quadratic form $\Lambda_{\mu\nu} r^\mu r^\nu$ in the entire $\RR^{N^2}$.
Clearly, it stays sign-definite and just increases in absolute value 
along any ray drawn from the origin.
Consider now the part of $\RR^{N^2}$ filled by rays where $\Lambda_{\mu\nu} r^\mu r^\nu <0$.
The potential is stable if and only if this ``negative region'' does not intersect the orbit space.

\begin{figure}[!htb]
   \centering
\includegraphics[width=7cm]{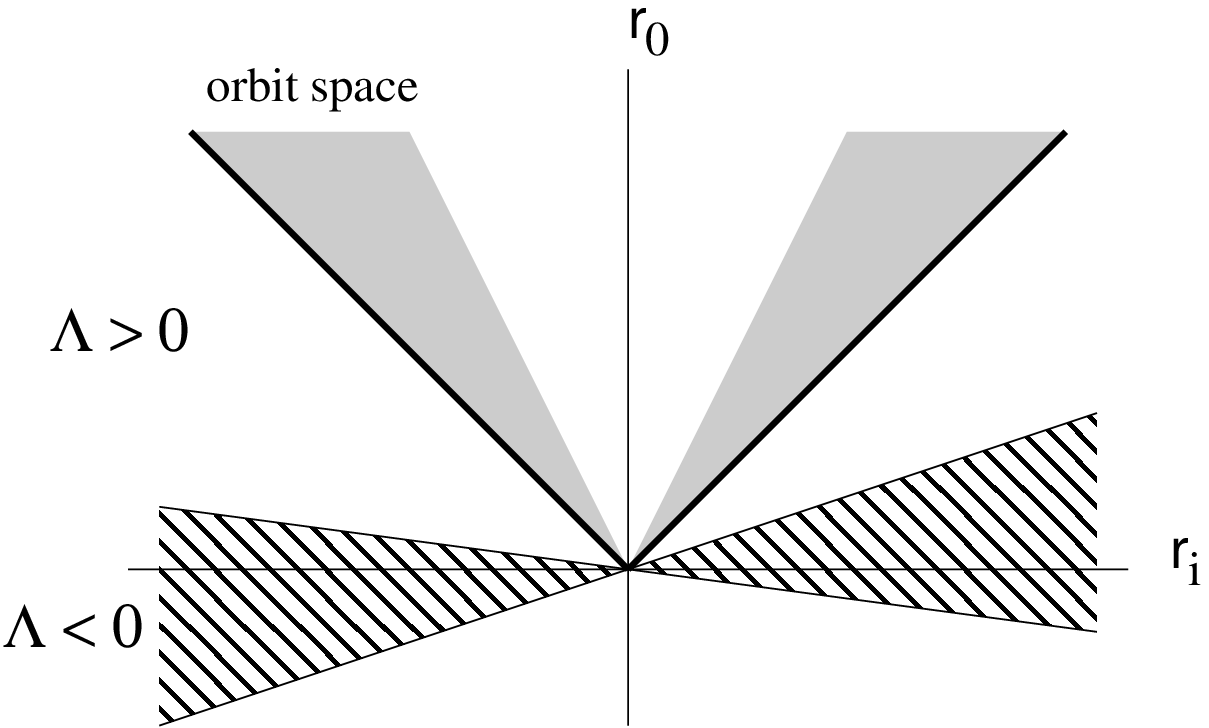}\hspace{2cm}
\includegraphics[width=7cm]{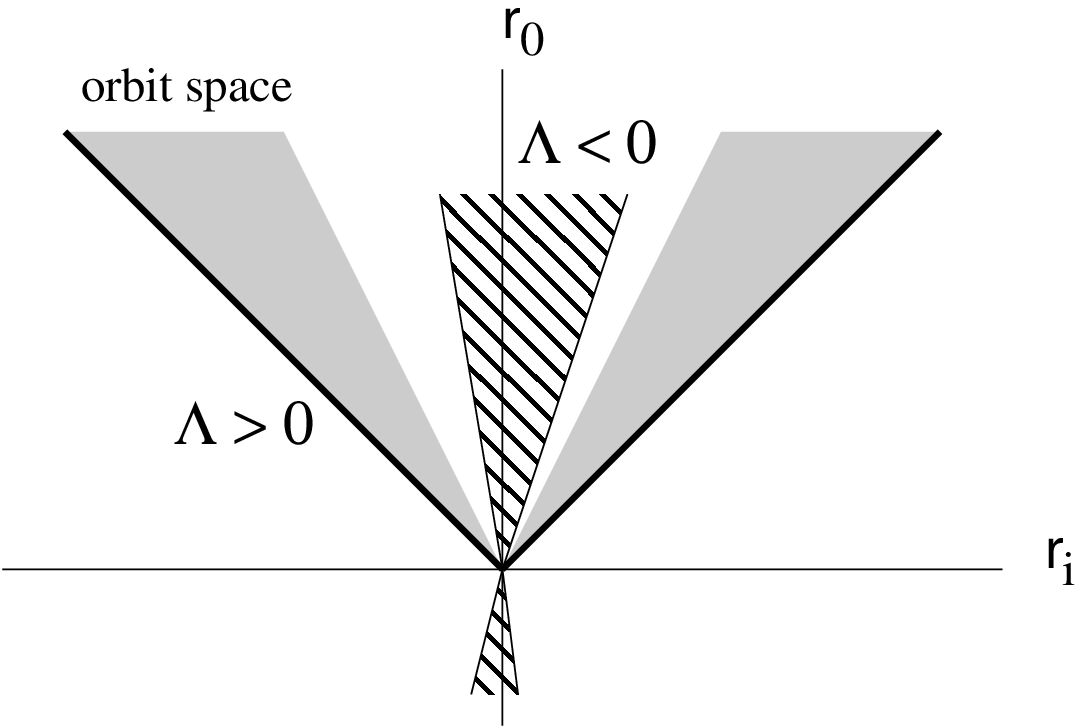}
\caption{Location of the negative region (shown as dashed area and labeled as $\Lambda < 0$) 
with respect to the orbit space on a generic $(r_0,r_i)$ plane.
The neutral orbit space is shown with thick lines, the charge-breaking orbit space is shown as a gray area. 
For 3HDM, the negative region can lie either outside the lightcone (left) or inside the empty inner cone (right).}
   \label{fig-positivity}
\end{figure}

This description is illustrated schematically in Fig.~\ref{fig-positivity} on a generic $(r_0,r_i)$-plane. 
Here, the neutral orbit space is indicated by the thick rays, 
the charge-breaking orbit space is shown by the gray region, while
the ``negative region'' is the dashed area labeled as $\Lambda < 0$. 
An interesting possibility absent in 2HDM is
that the ``negative region'' can be placed not only outside, but also inside the lightcone,
see Fig.~\ref{fig-positivity}, right.  
In this case the negative region is contained in the empty inner cone. An example of 3HDM potential
which realizes such an exotic situation is
\be
\label{potential:exotic}
V = - M_0 r_0 - M_i r_i + {1 \over 2}\Lambda_{00} r_0^2 + {1 \over 2}\Lambda_{ij} r_i r_j\,,
\ee
with a positive-definite $\Lambda_{ij}$ and a negative $\Lambda_{00}$ 
satisfying 
$$
0 > \Lambda_{00} > -{\Lambda_{i,min}\over 4}\,,
$$
where $\Lambda_{i,min}$ is the smallest eigenvalue of $\Lambda_{ij}$.
Note also that it is a specific property of 3HDM that the negative region can be placed either 
inside or outside the orbit space, because the orbit space of 3HDM isolates the inner cone from 
the region outside the lightcone. In NHDM with $N>3$ the orbit space is much more ``perforated'',
so that the inner cone is in fact connected  with the outer region.
Thus, intermediate situations between the two extremes shown in Fig.~\ref{fig-positivity} 
are also possible.

\subsection{Searching for the global minimum}

With the help of equipotential surfaces, the search for the global minimum of the potential 
can be reformulated in a very intuitive geometric way.

Suppose that we have made sure that the positivity conditions are satisfied. Then, an equipotential
surface ${\cal M}_C$ with a sufficiently negative $C$ does not intersect the orbit space.
Let us now increase the value of $C$. The lowest value of $C = C_0$, at which we have the first contact
with the orbit space, defines the {\em critical equipotential surface} ${\cal M}_{C_0}$. 
The contact point or points give the position(s) of the global minimum of the potential.
Thus, in order to identify the global minimum of a given potential, 
we need to find the unique equipotential surface that only touches but does not intersect
the orbit space. This fact, as well as the uniqueness of the critical equipotential surface, 
follow from the fact that the orbit space is a connected manifold and from
the nesting property of the equipotential surfaces. 

\begin{figure}[!htb]
   \centering
\includegraphics[width=9cm]{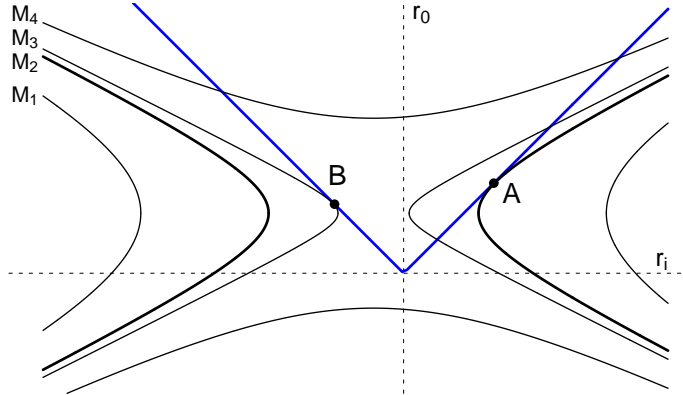}
\caption{Search for the global minimum via equipotential surfaces, see text for explanation.}
   \label{fig-equi1}
\end{figure}

This construction is illustrated in Fig.~\ref{fig-equi1} on a generic plane $(r_0, r_i)$. 
The outer boundary of the orbit space is shown schematically as a pair of rays.
Four equipotenial surfaces ${\cal M}_1$ through ${\cal M}_4$ of a Higgs potential are plotted as pairs of hyperboles.
The equipotential surface ${\cal M}_1$ corresponds to such a low value of $C$ 
that it never intersects the orbit space, so it is not realized in terms of doublets.
${\cal M}_2$ (shown in bold) is the lowest equipotential surface that touches the orbit space, 
its contact point $A$ giving the position of the global minimum. 
Any equipotential surface lying higher, such as ${\cal M}_3$ and ${\cal M}_4$, 
necessarily intersect the orbit space along some region, not just a single point.
However sometimes it can also lead to additional local minima, 
such as point $B$ on ${\cal M}_3$.

Although this geometric criterion for identification of the global minimum does not help 
to minimize a given potential,
it is of much use when we {\em construct} potentials with a specific explicit symmetry 
and a specific pattern of spontaneous symmetry breaking.
It will guarantees us that we are indeed working with the global minima of the potential.

\section{Symmetries of NHDM}\label{section:symmetries}

One of the main motivations to study non-minimal Higgs sectors is a possibility to introduce new symmetries
via the scalar sector. In particular, three questions are of much importance:
\begin{enumerate}
\item
Which symmetries can the lagrangian of a given non-minimal Higgs sector, e.g. NHDM, have?
\item
What are the possible ways to spontaneously break these symmetries? Which phenomena accompany symmetry breaking?
\item
How can these symmetries be extended into the fermion sector?
\end{enumerate}
A thoroughly studied example is again given by the 2HDM.
Various symmetries have been discussed starting from the original paper on the subject \cite{TDLee}, 
for a recent review see \cite{CPNSh}. 
Since various symmetries with the same group could be mapped to each other
by an appropriate reparametrization transformation, 
they represent the same symmetry of the Higgs lagrangian 
and might differ only in the fermionic sector.
Therefore, if one focuses on the scalar lagrangian only, one should look at 
conjugacy classes of symmetries, not at their particular implementations,
see \cite{ferreira-nhdm} for a discussion of this point. 

In \cite{ivanov2} the complete classifications of symmetry classes in the most general 2HDM was given.
Only six conjugacy classes of symmetry can exist in 2HDM, with groups $Z_2$, $(Z_2)^2$,
$(Z_2)^3$, $Z_2\times O(2)$, and $O(3)$. 
Reparametrization invariant criteria of the presence of a (hidden) symmetry were given in
\cite{haber2,heidelberg,nishi2006,ivanov0,ivanov1,ivanov2,nishi2008}, with a particular emphasis on
the $CP$-violation. 
Which of these symmetries can be extended into the fermionic sector has also received some attention
recently, see \cite{maximalCP,ferreira-generalizedCP,symbiotic}.

In NHDM, one can expect an even richer spectrum of possible symmetries and symmetry breaking;
some of them were studied in 
\cite{weinberg1976,Branco1980,DeshpandeHe1994,lavoura1994,Erdem1995,barroso2006,nishi2006,nishi:nhdm,ferreira-nhdm}. 
Nevertheless, despite all these efforts, we are still far from being able to write down the list
of all symmetries possible in NHDM.

In the Section we make steps towards this goal.

\subsection{Explicit symmetry group of NHDM}\label{section:explicit-symmetry}

An explicit symmetry of the Higgs lagrangian is a transformation
of the doublets $\phi_a$ that leaves the lagrangian invariant.
This is to be contrasted with the notion of reparametrization invariance of the model
mentione in the introduction, which refers to the invariance of physical observables
but not the parameters of the lagrangian itself.
In simple cases such symmetries are obvious and can be revealed directly 
from inspection of the Higgs potential. A more intricate situation takes place
when symmetries are hidden. They could become obvious after an appropriate reparametrization
transformation is applied, but such a transformation cannot be guessed in any simple way.

There are several issues concerning possible explicit symmetries of NHDM. 
In this paper we study the following questions:
\begin{itemize}
\item
Which explicit symmetry groups are allowed in NHDM?
\item
How can one find the explicit symmetry group of a given NHDM potential?
\item
How can one construct examples of NHDM realizing a given symmetry?
\end{itemize}
Here we make steps towards answering the first two questions, and the answer to the third one
will become clear on the way. 

We tackle the problem in the adjoint space $\RR^{N^2}$, which arguably 
gives the clearest view of possible symmetries of NHDM. 
Since we want the kinetic terms to be invariant, we consider here only (anti)unitary reparametrization transformations,
which induce proper rotations of $\vec r$, and keep $r_0$ and $\vec r^2$ unchanged.
Therefore, any symmetry of the potential is automatically a symmetry of the entire Higgs lagrangian. 

Let us first derive a general expression for the symmetry group of an NHDM potential.
We rewrite the Higgs potential (\ref{potential:again}) in the following form
\be
\label{V:rewritten}
V = - M_0 r_0 + {1\over 2}\Lambda_{00}r_0^2 + (-M_i + L_i r_0) r_i + {1\over 2}\Lambda_{ij}r_ir_j\,,\quad L_i \equiv \Lambda_{0i}.
\ee
Let us consider this expression in the entire adjoint space and study symmetry properties of
vectors $M_i$, $L_i$, and of the tensor $\Lambda_{ij}$ separately. In the discussion below 
we use short notation $d\equiv N^2-1$.

Let us denote by $G_M$ the subgroup of $O(d)$ that leaves the vector $M_i$ invariant.
If $M_i$ is non-zero, this group is given by all rotations and reflections in the $(d-1)$-dimensional space orthogonal 
to $M_i$: $G_M = O(d-1)$. If $M_i=0$, this group coincides with the entire $O(d)$.
The same is valid for $G_L$, the symmetry group of $L_i$. Obviously, if $M_i$ and $L_i$ are linearly independent,
the two $O(d-1)$ groups are different subgroups of $O(d)$ and intersect across $O(d-2)$.

Let us denote by $G_\Lambda$ the symmetry group of the tensor $\Lambda_{ij}$.
Since $\Lambda_{ij}$ is real and symmetric, it can always be diagonalized by an $SO(d)$ rotation,
and the symmetry group $G_\Lambda$ is then defined by the spectrum of $\Lambda_{ij}$.
If all $d$ eigenvalues are different, then $G_\Lambda = (Z_2)^{d}$.
If some of the eigenvalues coincide, then it is promoted to a continuous group according to the obvious rule:
each $k$-fold degenerate eigenvalue changes $(Z_2)^k \to O(k)$.
The generic representation is then 
\be
\label{G:Lambda}
G_\Lambda = O(k_1)\times O(k_2) \times \dots \times O(k_p)\times (Z_2)^{d - k_1 -\dots -k_p}\,.
\ee
Let us now describe the intersection of these three groups, $G_M \cap G_L \cap G_\Lambda$,
which gives the symmetry group of the quadratic form (\ref{V:rewritten}) in the entire space $\mathbb{R}^{d}$ of vectors $\vec r$.
This symmetry group consists of transformations, which leave invariant each of $M_i$, $L_i$, and $\Lambda_{ij}$.
This group is non-trivial if and only if there is an eigenvector of $\Lambda_{ij}$ orthogonal
to both $M_i$ and $L_i$. This statement can be also put in algebraic terms. Let us introduce the short notation
$\Lambda^na$ for a vector that can be obtained by acting of the $n$-th power of tensor $\Lambda_{ij}$ 
on a vector $a_i$. Then, let us denote the external product of $d$ vectors $a_i,\, b_i\, \dots\, z_i$ as
\be
(a,b,\dots,z)\equiv \epsilon_{i_1 i_2 \cdots i_d}\, a_{i_1} b_{i_2} \cdots z_{i_d}\,,
\ee
where $\epsilon_{i_1 i_2 \cdots i_d}$ is the standard fully-antisymmetric tensor in $d$ dimensions.
Finally, let us introduce the $r_0$-dependent vector $\mu_i(r_0) \equiv -M_i + L_i r_0$.
Then, the necessary and sufficient condition for existence of an eigenvector of $\Lambda_{ij}$ orthogonal
to both $M_i$ and $L_i$ is that the following polynomial in powers of $r_0$ is identically zero:
\be
\label{I-function}
{\cal I}(r_0) = (\mu,\Lambda \mu,\Lambda^2\mu, \cdots, \Lambda^{d-1}\mu) = 0\,.
\ee
If needed, this single criterion can be rewritten in terms of $M_i$ and $L_i$ separately.

If this criterion is satisfied, $G_M \cap G_L \cap G_\Lambda$ is a non-trivial group.
How large this group is depends on the {\em number} of eigenvectors of $\Lambda_{ij}$ orthogonal to both
$M_i$ and $L_i$ and on degeneracy of the corresponding eigenvalues. 
Each such eigenvector contributes a factor of $Z_2$ to the group,
and each $k$-tuple of such eigenvectors with degenerate eigenvalues gives rise to a factor of $O(k)$.
The resulting expression has the form of (\ref{G:Lambda}), with the total dimension $d$ replaced
by the dimension of the eigenspace orthogonal to $M_i$ and $L_i$.

The last step in identification of the symmetry group of NHDM is to take into account that not all
orthogonal transformations in $O(d)$ are realizable as reparametrization transformations of the doublets.
This group of physically realizable transformations, which we denote by $G_0$, coincides with the symmetry group of the 
orbit space. It consists of the reparametrization group $SU(N)$  
together with the corresponding anti-unitary transformations in the adjoint representation. 
With this notation, the symmetry group of NHDM can be finally written as
\be
\label{G:NHDM}
G = G_M \cap G_L \cap G_\Lambda \cap G_0\,.
\ee
Thus, the question of identification of all possible explicit symmetries of NHDM is reformulated
as the search for all possible types of intersections of the individual groups in (\ref{G:NHDM}).
We postpone the complete analysis for a future study.

We stress that even if identification of the symmetry group of a given Higgs potential
is difficult, the inverse task --- {\em construction} of potentials with desired symmetries --- is much easier.
Using the freedom in choosing $M_\mu$ and $\Lambda_{\mu\nu}$, we can align the vectors 
and the tensor with the orbit space in the way we like. 
The reader will find examples in Section~\ref{section:3HDM-examples}.

\subsection{Which discrete groups are realizable?}\label{section-realizable}

Additional discrete symmetries play an important role in theories beyond the Standard Model. 
However, not all of them are equally interesting in the context of our analysis.
Indeed, some discrete group will shape the phenomenology of the model in a distinct, group-specific way.
Other discrete groups, for example, any discrete subgroup that arises as a subgroup of 
a continuous symmetry group, 
do not tell us about the Higgs potential anything specific to this group
(see an extensive discussion of this point in \cite{ferreira-nhdm}).

The distinction is made clear by the following definition:
we call a discrete group $H$ {\em realizable} (in NHDM) if there exists an NHDM Higgs potential  
symmetric under $H$, but not symmetric under any continuous group containing $H$.  
For example, the NHDM potential with cubic symmetry,
\be
V = - m^2 \sum_{a=1}^N(\phi_a^\dagger \phi_a) + {\lambda \over 2} \sum_{a=1}^N(\phi_a^\dagger \phi_a)^2\,,
\ee
contains the symmetry group $S_N$ of permutations of the doublets, which is not a subgroup of any continuous symmetry
group of this potential. Thus, $S_N$ is realizable in NHDM.
On the other hand, the same potential is also invariant under phase shifts of the first doublet by multiples of $2\pi/q$,
for any integer $q$, giving a specific realization of the $Z_q$ symmetry group. 
However, this group is not realizable because it arises not on its own, 
but as a subgroup of the full $U(1)$ symmetry group of arbitrary phase shifts of the first doublet.

A natural question is then: which discrete groups are realizable in NHDM?
Note that the groups entering (\ref{G:NHDM}) contain either continuous groups or factors of $Z_2$
(which makes $Z_2$ a realizable group).
Nevertheless, we have just given an example of a realizable discrete group different from $(Z_2)^k$. 
Clearly, such symmetry groups must arise from non-trivial intersections of continuous groups.

Let us illustrate this point with a specific example.
In 3HDM, the field space is invariant under cyclic permutation of the three doublets
$\phi_1 \to \phi_2 \to \phi_3 \to \phi_1$.
However, this $Z_3$ group is a subgroup of the $U(1)$ group generated by $-\lambda_2 + \lambda_5 - \lambda_7$.
The corresponding finite-angle rotations of the doublets, $\tilde \phi_a = R_{ab}\phi_b$, are 
given by the matrix
\be
\label{z3continuous}
R_\alpha = 
{1\over 3}\left(\begin{array}{ccc} 
1+2\cos\alpha & 1+2\cos\alpha'' & 1+2\cos\alpha' \\ 
1+2\cos\alpha' & 1+2\cos\alpha & 1+2\cos\alpha'' \\ 
1+2\cos\alpha'' & 1+2\cos\alpha' & 1+2\cos\alpha 
\end{array}\right)\,,\quad \alpha' = \alpha + {2\pi \over 3}\,,\quad  \alpha'' = \alpha + {4\pi \over 3}\,.
\ee
The three elements of the $Z_3$ subgroup correspond to $\alpha = 0$, $2\pi/3$ and $4\pi/3$:
\be
\label{z3subgroup}
R_0 = 
\left(\begin{array}{ccc} 
1 & 0 & 0 \\ 
0 & 1 & 0 \\ 
0 & 0 & 1 \end{array}\right)\,,\quad 
R_{{2\pi \over 3}} = 
\left(\begin{array}{ccc} 
0 & 1 & 0 \\ 
0 & 0 & 1 \\ 
1 & 0 & 0 \end{array}\right)\,,\quad 
R_{{4\pi \over 3}} = 
\left(\begin{array}{ccc} 
0 & 0 & 1 \\ 
1 & 0 & 0 \\ 
0 & 1 & 0 \end{array}\right)\,.
\ee
In the adjoint space, the transformation (\ref{z3continuous}) with a generic $\alpha$ involves all eight coordinates $r_i$.
There are two invariant subspaces of vectors $\vec r$ under the action of this group: $E_{13468}= (r_1,r_3,r_4,r_6,r_8)$ 
and $E_{257}= (r_2,r_5,r_7)$. 
However, at $\alpha=2\pi n/3$, the first invariant subspace can be split into two smaller
invariant subspaces: the root plane $E_{38}=(r_3,r_8)$ and its complement $E_{146}= (r_1,r_4,r_6)$.
In fact, upon the action of so-defined $Z_3$ group we have precisely the cyclic permutations among
the ``usual'', ``primed'' and ``double-primed'' sets of coordinates discussed in Section~\ref{section:orbit}.

Let us now consider a Higgs potential with a maximal orthogonal symmetry in each of these subspaces, $E_{38}$, $E_{146}$ and $E_{257}$,
i.e. with symmetry group $G_M \cap G_L \cap G_\Lambda = G_\Lambda = O(2)\times O(3) \times O(3)$:
\be
\label{z3potential}
V= - M_0 r_0 + {1 \over 2}\lambda_{0}r_0^2 + {1 \over 2}\lambda_{38}(r_3^2+r_8^2) + {1 \over 2}\lambda_{146}(r_1^2+r_4^2+r_6^2)
+ {1 \over 2}\lambda_{257}(r_2^2+r_5^2+r_7^2)\,,
\ee
with suitable values of $M_0$ and $\lambda$'s. Here we chose $M_i = 0$ and 
$$
\Lambda_{\mu\nu} = \mathrm{diag}(\lambda_0,\,\lambda_{146},\,\lambda_{257},\,\lambda_{38},
\lambda_{146},\,\lambda_{257},\,\lambda_{146},\,\lambda_{257},\,\lambda_{38})\,.
$$
Due to $\lambda_{38}\not = \lambda_{146}$, 
this potential is not symmetric under the entire $U(1)$ group (\ref{z3continuous}), 
but it is symmetric under the discrete $Z_3$ subgroup (\ref{z3subgroup}). Thus, this subgroup arises from intersection
of the symmetry group of the orbit space, $G_0$, and the symmetry group of the potential, which in this case is given just by $G_\Lambda$.

With this example, it becomes clear how to identify a realizable discrete group in NHDM (other than powers of $Z_2$):
we just need to find a discrete group $H$ whose invariant-subspace decomposition of the $\vec r$-space 
differs from the $G$-invariant-subspace decomposition for any Lie group $G\subseteq G_0$ that contains $H$.

Indeed, let us pick up a Lie group $G \subseteq G_0$, which contains a discrete group $H$, and 
let the $G$-invariant-space decomposition of the $\vec r$-space be $\oplus E_i$. 
Clearly, the $H$-invariant-space decomposition, $\oplus E_{i,\alpha}$, can be only finer, not coarser; 
that is, it can be obtained form $\oplus E_i$ by either keeping each $E_i$ intact or
splitting it further into smaller subspaces $E_i \to \oplus_\alpha E_{i,\alpha}$. 
Now, recall that we have full freedom when constructing the Higgs potential. In particular,
we can construct a Higgs potential whose symmetry group is exactly $\otimes O(k_{i,\alpha})$, where each factor $O(k_{i,\alpha})$
is the symmetry group of rotations inside the $H$-invariant space $E_{i,\alpha}$ of dimension $k_{i,\alpha}$.  
If the $H$- and $G$-invariant-space decompositions are not identical, then the potential is not symmetric under the full group $G$.
If this situation holds for all possible Lie groups $G \subseteq G_0$, then the discrete symmetry of the potential under $H$ cannot
be extended into any continuous one, thus, the discrete group $H$ is realizable.

This strategy can be applied to NHDM to generate the full list of discrete groups realizable at given $N$.
We postpone this analysis for a future work.

\subsection{Spontaneous symmetry breaking}

Even if the Higgs lagrangian has a non-trivial explicit symmetry, this symmetry can be spontaneously 
broken in the vacuum state.  In other words, the symmetry group in the vacuum, $G_v$,
can be smaller than the explicit symmetry group $G$ of the Higgs potential.

This possibility brings in several natural questions:
\begin{itemize}
\item
How exactly can each of the explicit symmetries be spontaneously broken?
\item
How can one recognize at the level of potential that an explicit symmetry is spontaneously broken?
\item
How can one construct examples of NHDM realizing a given pattern of spontaneous symmetry breaking?
\end{itemize}
Let us first give a general expression for $G_v$. 
In the derivation of the explicit symmetry group $G$ given by Eq.~\ref{G:NHDM}, we searched for all possible
orthogonal transformations of the orbit space that left the potential invariant. Now, deriving $G_v$, we require that only 
those transformations that leave the v.e.v.'s intact are allowed. In other words, we must replace $G_0$ by the isotropy group
of $\lr{\vec r}$, which we denote by $G_{\lr{\vec r}}$:  
\be
\label{Gv:NHDM}
G_v = G_M \cap G_L \cap G_\Lambda \cap G_{\lr{\vec r}}\,.
\ee
Using the isotropy groups for the charge-breaking and neutral vacua found in \cite{NHDM2010},
one can in principle write down the full list of spontaneous symmetry breaking patterns possible in NHDM.
We postpone this study, too, for a future work.

If we know that the Higgs potential has a non-trivial explicit symmetry, i.e. (\ref{I-function}) is satisfied,
and if we have an expression for the position of the global minimum $\lr{r^\mu}$, then we can check whether 
the vacuum violates this symmetry. 
Geometrically, the criterion is whether the vector $\lr{r_i}$ is orthogonal to the same eigenvectors of $\Lambda_{ij}$
as $M_i$ and $L_i$. If this does not hold for at least one eigenvector,
then the symmetry is spontaneously broken.

These observations are straightforward, but they do not resolve the problem of detecting spontaneous symmetry breaking 
at the level of potential (that is, they require the knowledge of $\lr{r^\mu}$). 
In principle, if we know the potential, then all the properties of the vacuum 
including the possible spontaneous symmetry breaking must follow.
There must exist a condition involving only $\Lambda_{\mu\nu}$ and $M_\mu$ for spontaneous breaking of a given symmetry.
For example, in 2HDM this condition for spontaneous breaking of $CP$-violation was derived in \cite{ivanov1,ivanov2}
and was based on non-trivial geometric properties of the potential. It remains to be seen whether a similar derivation
can be conducted in a general NHDM.

\section{Examples of 3HDM with specific symmetry patterns}\label{section:3HDM-examples}

\subsection{Decoupling of directions}

In this Section we will make use of the general results of the two previous Sections and construct 
several 3HDM potentials, which exhibit various patterns of discrete symmetry breaking.
The strategy we use is the following. In each case we first consider a lower-dimensional section of the orbit space,
and then draw a quadric (the critical equipotential surface) with desired symmetry
that touches the orbit space at isolated points.
Then we write down a Higgs potential that corresponds to such an equipotential surface;
the symmetries of the Higgs potential and of the vacuum states will follow.

Two issues in this procedure require justification.
First, let us reiterate the point that when constructing the Higgs potential we have the full freedom in choosing its parameters, 
provided the positivity conditions are satisfied. It means that any quadric, 
no matter where it is placed and how it is oriented with respect to
the orbit space, is an equipotential surface of a suitable Higgs potential.

\begin{figure}[!htb]
   \centering
\includegraphics[width=8cm]{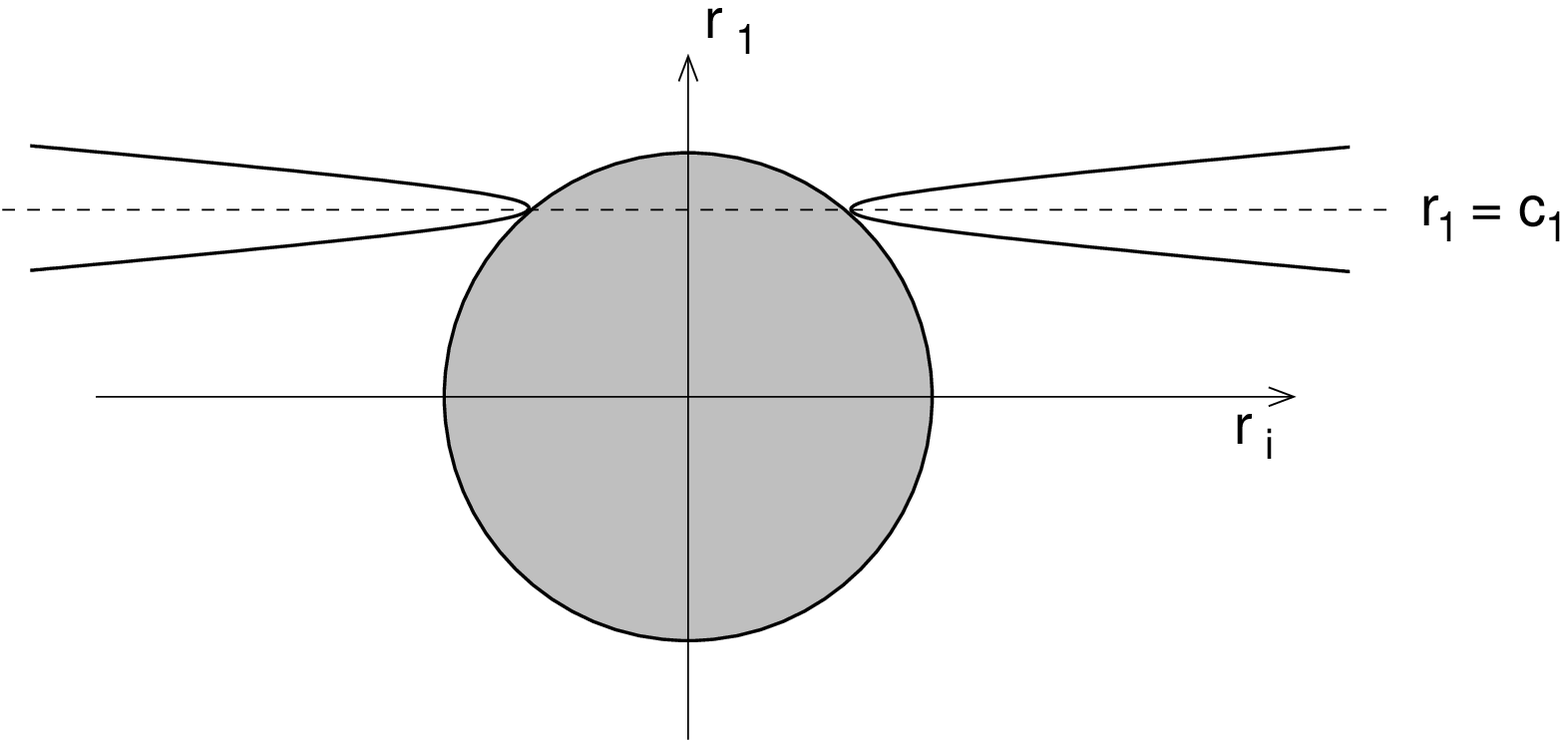}\hspace{2cm}
\includegraphics[width=5cm]{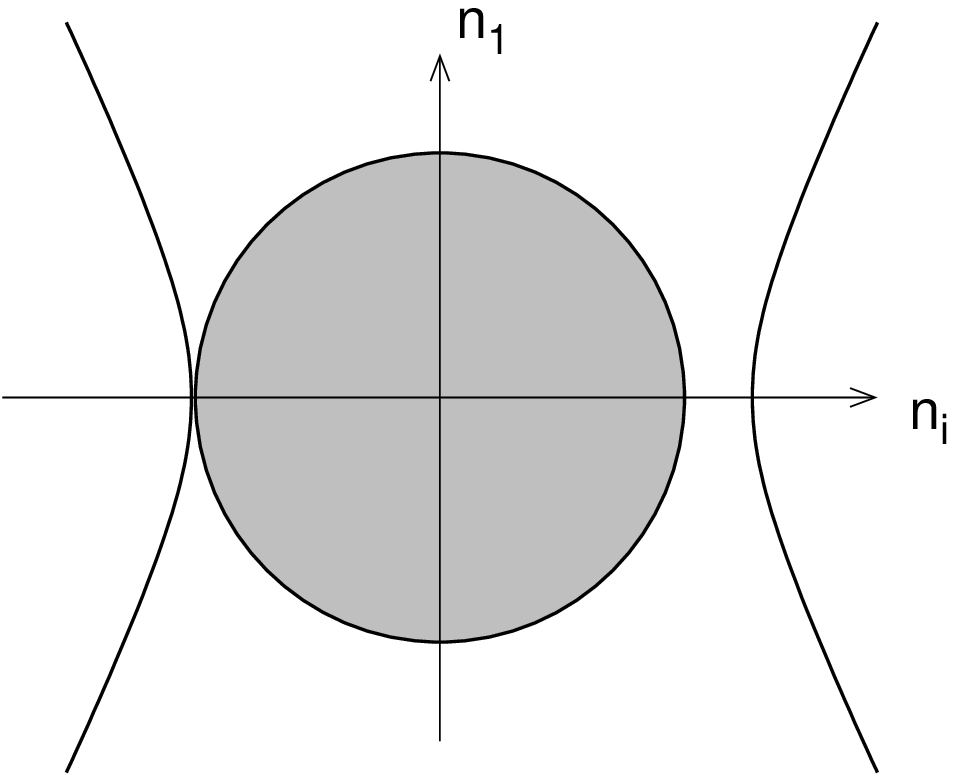}
\caption{Illustration of decoupling of unwanted directions, see explanations in the main text. 
The gray circle shows schematically the orbit space, the hyperboles are the critical equipotential surfaces.}
   \label{fig-decoupling}
\end{figure}

The second point is that when constructing specific cases we can safely focus on any affine subspace 
of the adjoint space $\mathbb{R}^{N^2}$ without worrying what happens outside it.
In order to illustrate this procedure of ``decoupling'' of unwanted directions,
let us choose a diagonal and non-singular $\Lambda_{\mu\nu}$. The linear term in the potential 
can be compensated by a shift of $(r_0,r_i)$, and up to the overall constant we have
\be
\label{V:diagonal}
V = {1\over 2}\lambda_0 (r_0-c_0)^2 + {1\over 2}\sum_i \lambda_i (r_i-c_i)^2\,. 
\ee
Here, $c_0 = M_0/\lambda_0$, $c_i = M_i/\lambda_i$. All $c$'s and $\lambda$'s can be chosen independently.
Suppose that $\lambda_0>0$, and one of $\lambda_i$, for example, $\lambda_1$, is positive 
and very large: $\lambda_1 \gg \lambda_0, |\lambda_i|$, $i \not = 1$.  
Then, the equipotential surfaces are very oblate in the direction of $r_1$, see Fig.~\ref{fig-decoupling}, left.
An equipotential surface ${\cal M}_C$ with a sufficiently small $C$
approaches the orbit space most closely at $r_1 \to c_1$ and quickly runs away as $r_1$ deviates from $c_1$.
In the limit $\lambda_1 \to \infty$, all the contact points between the equipotential surface and the orbit space, 
which define the minima of the potential, are restricted to the subspace $r_1 = c_1$.
So, one can focus on this subspace and disregard the fact that the orbit space extends beyond it.
In this way several directions can be ``decoupled'' from the analysis.

This procedure can be called the ``strong asymptotic decoupling'' of unwanted axes. 
It works for any contact --- symmetric or asymmetric, from inside or from outside --- 
between an equipotential surface and the orbit space.
However, in certain cases a relaxed and more useful version of decoupling can be used.
For example, let us consider a constant-$r_0$ section of the orbit space.
The lightcone condition restricts the orbit space to lie in a region
on and inside the unit ball $\sum_i n_i^2 \le 1$.
Let us again pick up axis 1 and consider various $n_1=\mathrm{const}$ sections of the orbit space.
The orbit space has the largest extent along $n_i$, $i\not = 1$, 
at $n_1 = 0$, which is very schematically illustrated by Fig.~\ref{fig-decoupling}, right.
Therefore, if $c_1 = 0$ in (\ref{V:diagonal}), if it sufficient to take any non-negative 
$\lambda_1$ to safely decouple axis 1 and guarantee that the global minimum is located at $n_1=0$. This includes the 
``cylindric case'', $\lambda_1=0$, i.e. when the potential does not depend on the coordinate $n_1$ at all.

\subsection{Spontaneously broken $Z_3$: neutral vacuum}

We now focus on 3HDM and start with the $Z_3$ symmetry group 
generated by cyclic permutation of the three doublets, 
discussed in Section~\ref{section-realizable}. 
A Higgs potential explicitly realizing this symmetry is:
\be
V = - M_0 r_0 + {1\over 2}\lambda_0 r_0^2 + {1\over 2}\lambda_{38} (r_3^2 + r_8^2)
+ {1\over 2}\lambda_\perp (|r_{12}|^2 + |r_{45}|^2 + |r_{67}|^2)\,.\label{Z3explicit}
\ee
In terms of doublets, the potential has the form
\bea
V &=& - {M_0 \over \sqrt{3}}[(\fd_1\phi_1)+(\fd_2\phi_2)+(\fd_3\phi_3)] + 
{\lambda_0+\lambda_{38} \over 6} \left[(\fd_1\phi_1)^2+(\fd_2\phi_2)^2+(\fd_3\phi_3)^2\right]\nonumber\\[2mm]
&&+\ {2\lambda_0-\lambda_{38} \over 6}[(\fd_1\phi_1)(\fd_2\phi_2)+(\fd_2\phi_2)(\fd_3\phi_3)+(\fd_3\phi_3)(\fd_1\phi_1)]\nonumber\\
&&+\ {\lambda_\perp \over 2}\left[|\fd_1\phi_2|^2+|\fd_2\phi_3|^2+|\fd_3\phi_1|^2\right]\nonumber\,.
\eea
Note that in principle we could add terms linear in $r_1+r_4+r_6$ and $r_2+r_5+r_7$, which are also $Z_3$-symmetric. 
However, we prefer to use the geometric argument just discussed to decouple all the ``transverse'' directions 
and to focus only on the root plane $(n_3,n_8)$.

\begin{figure}[!htb]
   \centering
\includegraphics[width=6cm]{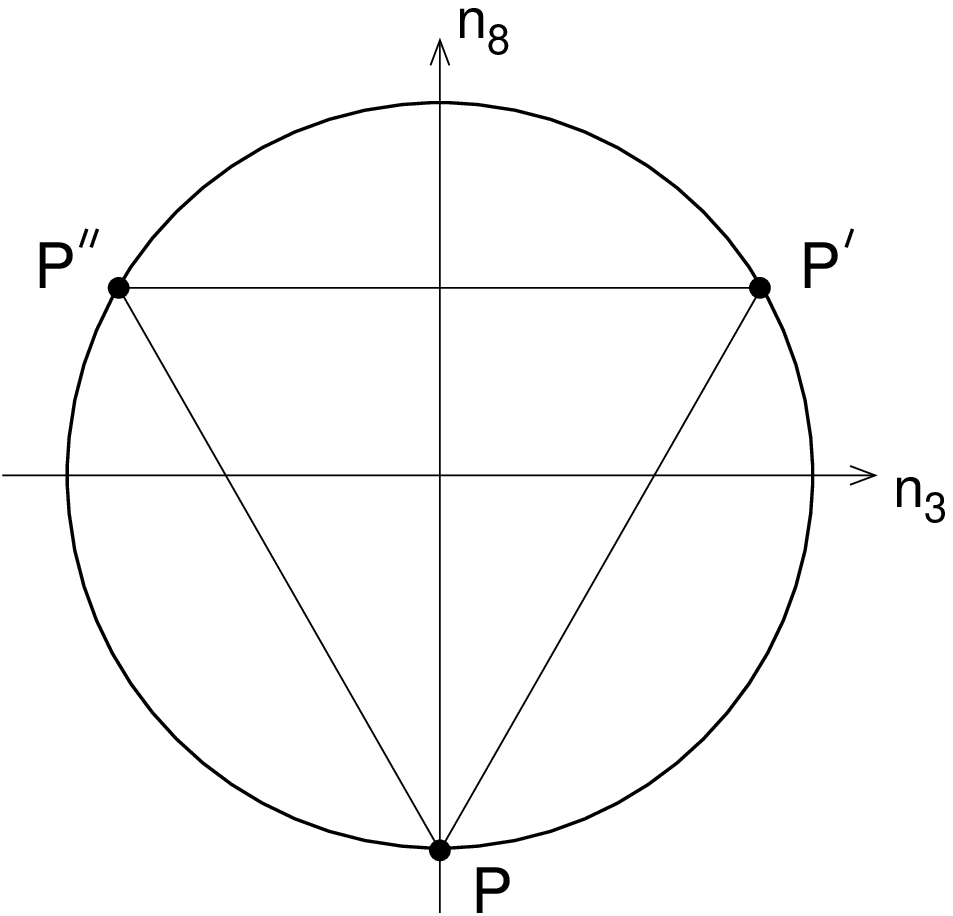}\hspace{3cm}
\includegraphics[width=6cm]{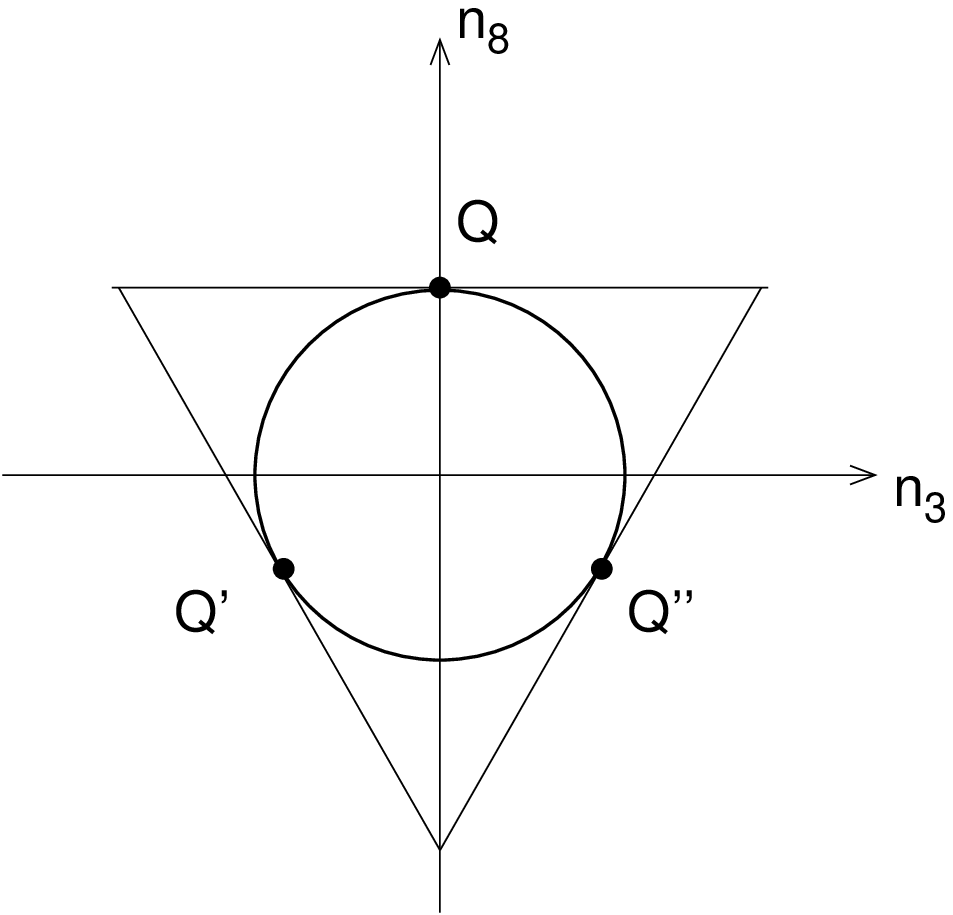}
\caption{Spontaneous $Z_3$-violation into neutral (left) and charge-breaking (right) minima. The critical equipotential surfaces 
are circles; the points of the global minimum of the potential are indicated by dots.}
   \label{fig-z3}
\end{figure}

To spontaneously break the $Z_3$-symmetry, the potential must have the global minimum at 
three points lying on the root plane:
\bea
&P:&  
\lr{\phi_1} = \doublet{0}{0}\,,\quad 
\lr{\phi_2} = \doublet{0}{0}\,,\quad 
\lr{\phi_3} = {1\over\sqrt{2}}\doublet{0}{v}\,,\nonumber\\[3mm]
&P'':&  
\lr{\phi_1} = \doublet{0}{0}\,,\quad 
\lr{\phi_2} = {1\over\sqrt{2}}\doublet{0}{v}\,,\quad
\lr{\phi_3} = \doublet{0}{0}\,,\nonumber\\[3mm]
&P':&  
\lr{\phi_3} = {1\over\sqrt{2}}\doublet{0}{v}\,,\quad 
\lr{\phi_1} = \doublet{0}{0}\,,\quad 
\lr{\phi_2} = \doublet{0}{0}\,.\nonumber
\eea
The critical equipotential surface is therefore the circumscribed circle of the triangle, see Fig.~\ref{fig-z3}, left. 
In order for these points to be minima, it is sufficient to take any $\lambda_\perp > 0$, together with
positive $\lambda_0$ and negative $\lambda_{38}$, provided that $\lambda_0 + \lambda_{38} > 0$. 
In the entire $r^\mu$ space, the critical equipotential surface is a hyperboloid that touches the forward lightcone
along the circle $r_3^2+r_8^2=r_0^2$ at some $r_0$.

Expanding the potential at a minimum point up to quadratic terms, one can find the Higgs boson masses:
\be
m_h^2 = {v^2\over 3}(\lambda_0 + \lambda_{38})\,,\quad m_{H^\pm}^2 = {v^2 \over 4}(-\lambda_{38})\,,\quad 
m_{H^0}^2 = {v^2 \over 4}(\lambda_\perp - \lambda_{38})\,.
\ee
Here $h$ is the Higgs boson along the v.e.v., $H^0$ are all the remaining neutral Higgs bosons,
and $H^\pm$ are all the charged Higgs bosons.

Note that in addition to the discussed $Z_3$-symmetry, this potential has other explicit symmetries, both discrete and continuous. 
We do not discuss them here because the sole purpose of this example was to illustrate
the $Z_3$ breaking in the computationally simplest form. 

\subsection{Spontaneously broken $Z_3$: charge-breaking vacuum}

Consider now the same $Z_3$-symmetric potential (\ref{Z3explicit}), but with $\lambda_{38} >0$.
In this case the equipotential surfaces in the $r^\mu$-space are concentric ellipsoids sitting inside the ``hole'' 
of the orbit space.
The critical equipotential surface is the ellipsoid that touches the orbit space from inside.
If $\lambda_\perp$ is large enough, then the contact points are again located in the root plane.
The $(n_3,n_8)$ section of the critical equipotential surface is represented on this plane
by the circle of radius 1/2, see Fig.~\ref{fig-z3}, right.

This potential has the global minima at the three maximally charge-breaking points:
\bea
&Q:&  
\lr{\phi_1} = {1\over \sqrt{2}}\doublet{0}{v}\,,\quad 
\lr{\phi_2} = {1\over \sqrt{2}}\doublet{v}{0}\,,\quad 
\lr{\phi_3} = \doublet{0}{0}\,,\nonumber\\[3mm]
&Q'':&  
\lr{\phi_1} = {1\over \sqrt{2}}\doublet{v}{0}\,,\quad 
\lr{\phi_2} = \doublet{0}{0}\,,\quad
\lr{\phi_3} = {1\over \sqrt{2}}\doublet{0}{v}\,,\nonumber\\[3mm]
&Q':&  
\lr{\phi_3} = \doublet{0}{0}\,,\quad 
\lr{\phi_1} = {1\over \sqrt{2}}\doublet{0}{v}\,,\quad 
\lr{\phi_2} = {1\over \sqrt{2}}\doublet{v}{0}\,.\nonumber
\eea
An explicit calculation gives the following Higgs masses:
\be
m_i^2 = {\lambda_\perp \over 2}v^2\,,\quad {\lambda_\perp - \lambda_{38} \over 4}v^2\,,\quad  {\lambda_{38} \over 2}v^2\,,\quad
{4\lambda_0 + \lambda_{38} \over 12}v^2\,. 
\ee 
The positivity of all the masses squared are guaranteed by $\lambda_\perp > \lambda_{38} > 0$.

\subsection{Other symmetry patterns on the root plane}

\begin{figure}[!htb]
   \centering
\includegraphics[width=6cm]{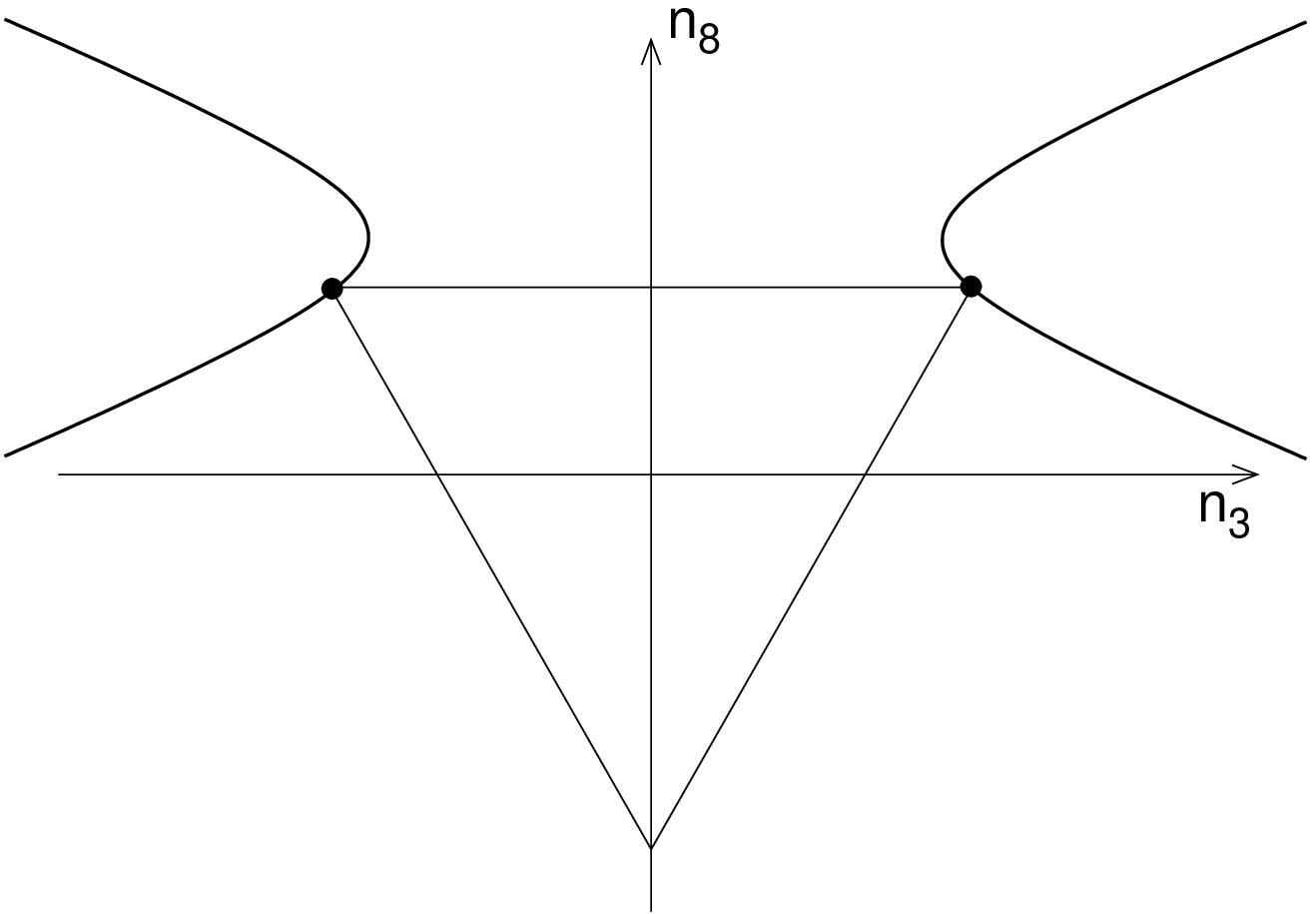}\hspace{2cm}
\includegraphics[width=6cm]{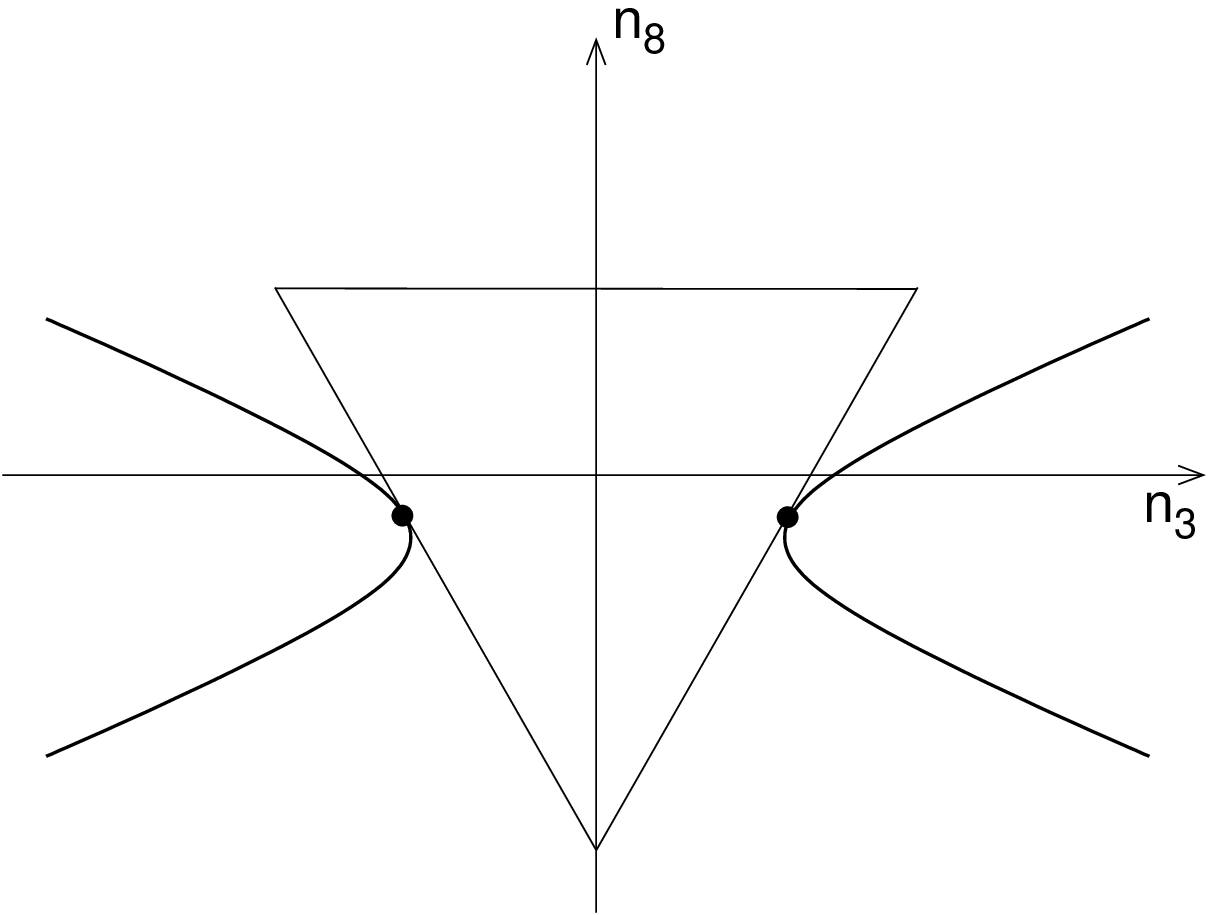}
\caption{Spontaneous $Z_2$-violation into neutral (left) and charge-breaking (right) minima. 
Hyperboles are the critical equipotential surfaces, the positions of the minima are indicated by dots.}
   \label{fig-z2}
\end{figure}

\begin{figure}[!htb]
   \centering
\includegraphics[width=6cm]{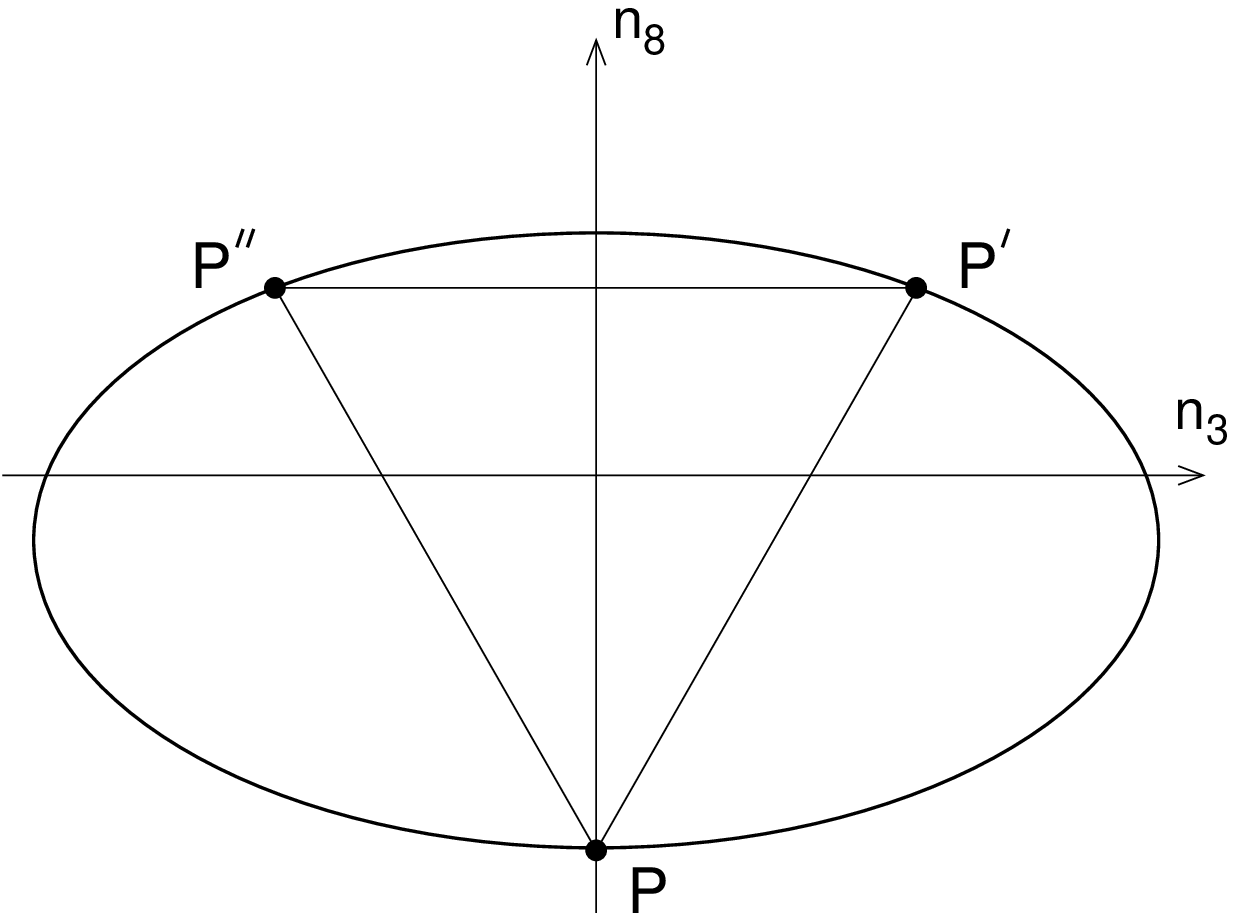}\hspace{2cm}
\includegraphics[width=6cm]{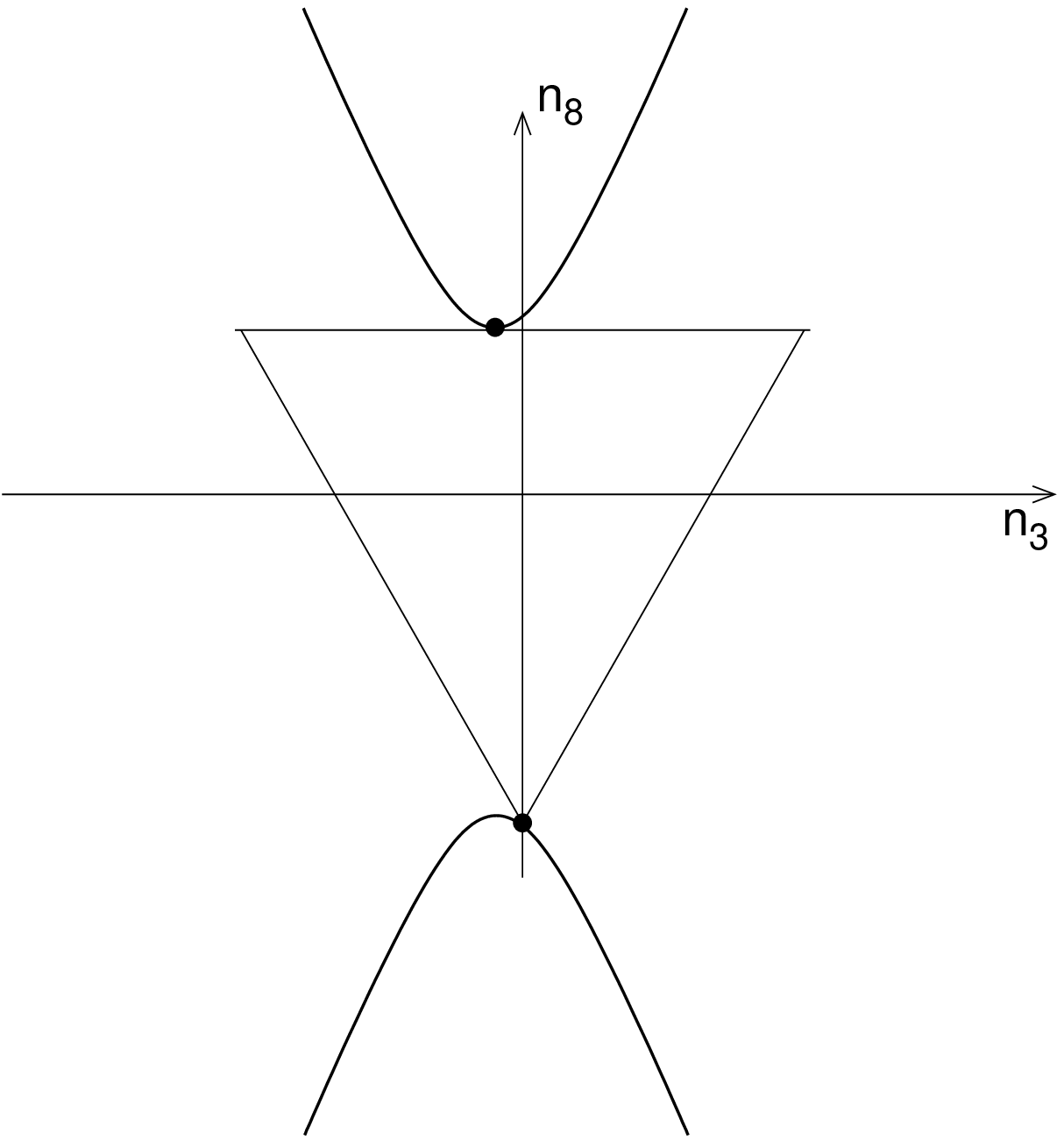}
\caption{Coexistence of degenerate minima with different symmetries. Left: coexistence
of $Z_2$-symmetric and $Z_2$-violating minima. Right: coexistence of charge-breaking and neutral minima.}
   \label{fig-coexist}
\end{figure}

Other patterns of symmetry breaking with minima localized on the root plane
can be devised with Higgs potentials having less symmetry in the root plane than (\ref{Z3explicit}). 
This can be achieved, for example, by replacing
\be
\lambda_{38}(r_3^2+r_8^2) \to \lambda_{3}r_3^2+ \lambda_{8}r_8^2\quad \mbox{with}\quad \lambda_3 \not = \lambda_8\,,
\ee
and by introducing $M_3, M_8 \not = 0$. The $(n_3,n_8)$ sections of the equipotential surfaces can then be
ellipses, hyperboles or parabolas, depending on the signs of $\lambda_3, \lambda_8$ (as usual, we keep $\lambda_0$ and $\lambda_\perp$ positive). 
Several examples are shown in Figures~\ref{fig-z2} and \ref{fig-coexist}.
\begin{itemize}
\item
$\lambda_3 < 0$, $\lambda_8 > 0$: spontaneous breaking of the $Z_2$ symmetry into neutral (Fig.~\ref{fig-z2}, left) 
or charge-breaking (Fig.~\ref{fig-z2}, right) vacuum.
\item
$\lambda_8 < \lambda_3 < 0$: triple degenerate neutral vacuum, Fig.~\ref{fig-coexist}, left, 
with coexisting points that break and conserve the $Z_2$ symmetry. A similar picture for charge-breaking vacua is also possible
for $\lambda_8 > \lambda_3 > 0$.
\item
$\lambda_3 > 0$, $\lambda_8 < 0$: double degenerate minimum with coexistence of neutral and charge-breaking vacua, 
Fig.~\ref{fig-coexist}, right. In this Figure we deliberately shifted the equipotential surface to illustrate
that the coexistence and degeneracy of neutral and charge-breaking vacua does not require any symmetry of the potential.
\end{itemize}
In all cases the values of $M_3$ and $M_8$ are assumed to be appropriately adjusted.

\subsection{Spontaneous breaking of $D_4$}

More complicated symmetry breaking patterns arise in the transverse space. 
For example, let us construct a Higgs potential with the dihedral symmetry group $D_4$, the symmetry group of a square.
A particular representation of this symmetry in the space of Higgs doublets is 
given by the group of sign flips of the first or of the second doublets as well as exchange
$\phi_1 \leftrightarrow \phi_2$ with $\phi_3$ always kept fixed.
In the adjoint space, these different kinds of $Z_2$ transformations correspond to:
\bea
\phi_1 \to -\phi_1 & \quad \Rightarrow \quad & n_{12} \to -n_{12}\,,\quad n_{45} \to -n_{45}\,,\nonumber\\
\phi_2 \to -\phi_2 & \quad \Rightarrow \quad & n_{12} \to -n_{12}\,,\quad n_{67} \to -n_{67}\,,\nonumber\\
\phi_1 \leftrightarrow \phi_2 & \quad \Rightarrow \quad & n_3 \to -n_3\,\quad n_2 \to -n_2\,\quad n_{45} \leftrightarrow n_{67}^*\,.
\eea
An example of a $D_4$-symmetric potential is
\bea
\label{D4potential}
V & = & - M_0 r_0 + {1 \over 2}\lambda_0 r_0^2 + {1 \over 2}\lambda_1 r_1^2 + {1 \over 2}\lambda_{46}(r_4^2+r_6^2) 
+ {1 \over 2}\lambda_{257} (r_2^2 + r_5^2 + r_7^2)\\
&=& - {M_0 \over \sqrt{3}}[(\fd_1\phi_1)+(\fd_2\phi_2)+(\fd_3\phi_3)] + 
{\lambda_0 \over 6} [(\fd_1\phi_1)+(\fd_2\phi_2)+(\fd_3\phi_3)]^2\nonumber\\
&&+\ {\lambda_1\over 2}[\Re(\fd_1\phi_2)]^2
+ {\lambda_{46}\over 2}\left\{[\Re(\fd_2\phi_3)]^2 + [\Re(\fd_3\phi_1)]^2\right\}\nonumber\\
&&+\ {\lambda_{257}\over 2}\left\{[\Im(\fd_1\phi_2)]^2 + [\Im(\fd_2\phi_3)]^2 + [\Im(\fd_3\phi_1)]^2\right\}\nonumber\,.
\eea
with positive $\lambda_0$, $\lambda_1$ and $\lambda_{257}$, and negative $\lambda_{46}$.
Using the decoupling arguments, we can be sure that the global minimum is located in the $n_1 = n_2 = n_5 = n_7=0$ subspace.
Then we note that the Higgs potential (\ref{D4potential}) does not depend on $n_3$ and $n_8$;
therefore, the equipotential surfaces are cylindric along these two directions.
Projecting them onto the $(n_4,n_6)$ plane, we find concentric circles around the origin because the potential
is $O(2)$-symmetric in this plane.

\begin{figure}[!htb]
   \centering
\includegraphics[width=6cm]{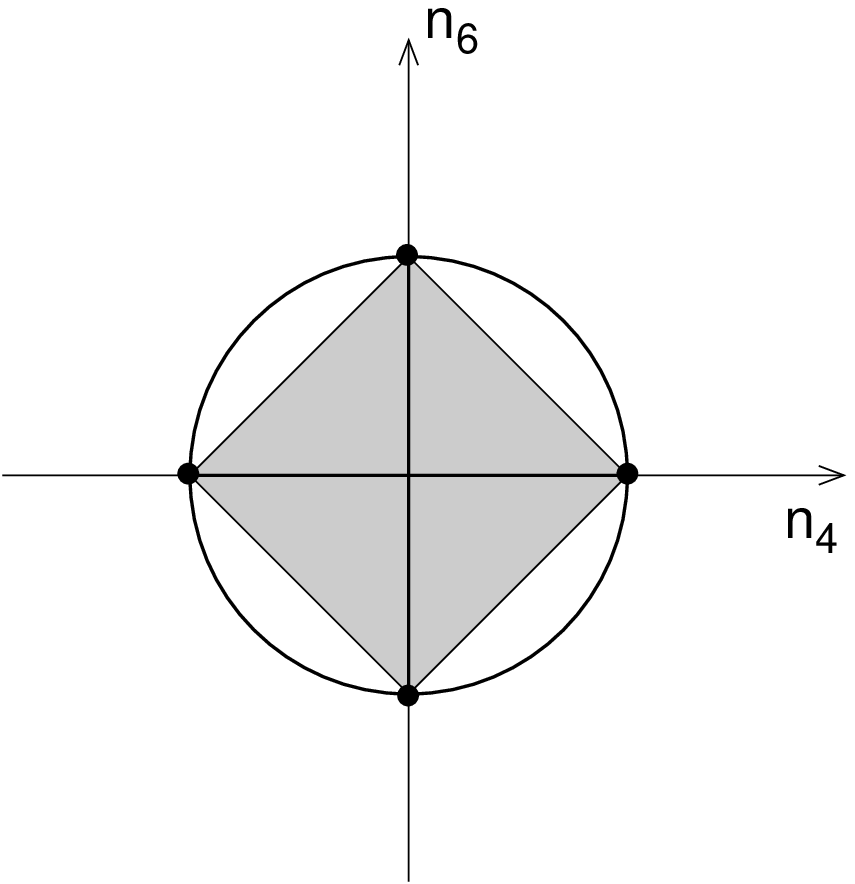}\hspace{2cm}
\includegraphics[width=6cm]{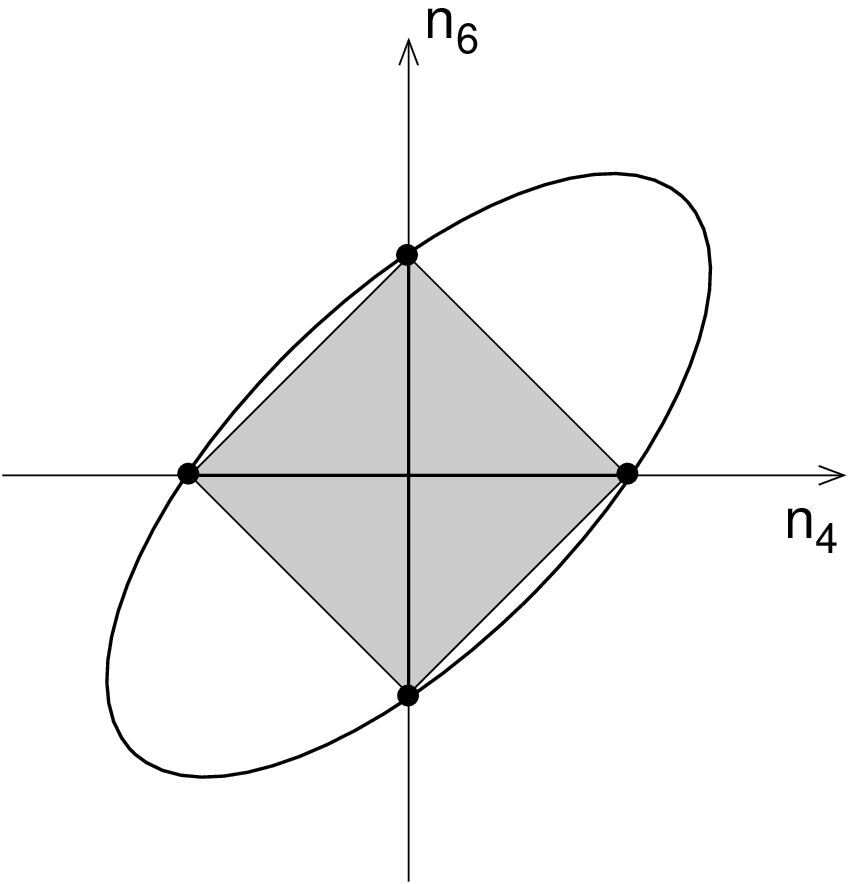}
\caption{Patterns of symmetry violation on the $(n_4, n_6)$-plane. Left: spontaneous violation of $D_4$ down to $Z_2$;
right: spontaneous violation of $Z_2\times Z_2$.}
   \label{fig-d4}
\end{figure}

In the Appendix we describe the geometry of the $n_1 = n_2 = n_5 = n_7=0$ section of the orbit space. 
When projected onto the $(n_4,n_6)$ plane, it has the square shape shown in Fig.~\ref{fig-d4}.
The four vertices correspond to neutral vacua with coordinates
\bea
&& n_4 = \pm {\sqrt{3} \over 2}\,,\quad n_6 = 0\,,\quad n_3 = - {\sqrt{3} \over 4}\,,\quad n_8 = -{1 \over 4}\,,\nonumber\\
&& n_4 = 0\,,\quad n_6 = \pm {\sqrt{3} \over 2}\,,\quad n_3 = + {\sqrt{3} \over 4}\,,\quad n_8 = -{1 \over 4}\,.
\eea
The v.e.v.'s of the lower components of the three doublets at these points are
\be
(\lr{\phi_1^0},\, \lr{\phi_2^0},\, \lr{\phi_3^0}) = {1\over \sqrt{2}}(0,\,\pm v,\,v)\quad \mbox{or} 
\quad {1\over \sqrt{2}}(\pm v,\,0,\, v)\,.
\ee
The critical equipotential surface of the potential (\ref{D4potential}) is the circle that goes through
the four vertices of the square, Fig.~\ref{fig-d4}, left. One gets four degenerate global minima, 
and at each minimum the $D_4$ symmetry group is broken to $Z_2$ group of sign flips of the doublet with zero v.e.v.
Thus, we have an example of spontaneous breaking $D_4 \to Z_2$.
An explicit expansion of the Higgs potential at the minimum gives the following physical Higgs boson masses:
\bea
h_i^\pm: && m^2_i = -{\lambda_{46} \over 2}v^2\,,\quad  -{\lambda_{46} \over 4}v^2\,,\\
h^0_i: && m^2_i = {\lambda_1 \over 4}v^2\,,\quad {2\lambda_{257} - \lambda_{46} \over 4}v^2\,,
\quad -{\lambda_{46} \over 2}v^2\,,\quad {4\lambda_0 + 3\lambda_{46} \over 6}v^2\,,
\quad {4\lambda_{257} - \lambda_{46} \over 2}v^2\,.\nonumber
\eea
All these masses squared are positive once $0 > \lambda_{46} > -4\lambda_0/3$.

One can also think of a less symmetric potential in the $(n_4,n_6)$-plane.
For example, by replacing 
\be
\lambda_{46}(r_4^2+r_6^2) \to \lambda_{46+}(r_4+r_6)^2 + \lambda_{46-}(r_4-r_6)^2\,,
\ee
with $\lambda_{46+} \not = \lambda_{46-}$ but both being negative,
we get a potential symmetric under the group $Z_2\times Z_2$ generated by $\phi_1 \leftrightarrow \phi_2$ and
by simultaneous sign flips of $\phi_1$ and $\phi_2$ (with $\phi_3$ kept constant).
Its equipotential surfaces on the $(n_4,n_6)$ are ellipses elongated along the $n_4 + n_6 = 0$ or 
$n_4 - n_6 = 0$ directions.
The critical equipotential surface will again touch the square at four vertices as shown in Fig~\ref{fig-d4}, right.
The $Z_2 \times Z_2$ symmetry is broken completely at each of the four degenerate minima.

\subsection{Spontaneous breaking of the octahedral symmetry}

Other symmetry groups can be realized in higher-dimensional subspaces. 
Let us consider, for example, the following potential
\bea
\label{octapotential}
V &=& - M_0 r_0 + {1 \over 2}\lambda_0 r_0^2 + {1 \over 2}\lambda_{246}(r_2^2 + r_4^2 +r_6^2)
+ {1 \over 2}\lambda_{157} (r_1^2 + r_5^2 + r_7^2)\\
&=& -\ {M_0 \over \sqrt{3}}[(\fd_1\phi_1)+(\fd_2\phi_2)+(\fd_3\phi_3)] + 
{\lambda_0 \over 6} [(\fd_1\phi_1)+(\fd_2\phi_2)+(\fd_3\phi_3)]^2\nonumber\\
&&+\ {\lambda_{246}\over 2}\left\{[\Im(\fd_1\phi_2)]^2 + [\Re(\fd_2\phi_3)]^2 + [\Re(\fd_3\phi_1)]^2\right\}\nonumber\\
&&+\ {\lambda_{157}\over 2}\left\{[\Re(\fd_1\phi_2)]^2 + [\Im(\fd_2\phi_3)]^2 + [\Im(\fd_3\phi_1)]^2\right\}\nonumber\,,
\eea
with positive $\lambda_0$, $\lambda_{157}$ and negative $\lambda_{246}$. 
The mismatch between the real and imaginary parts of the doublets' cross products is introduced here on purpose.
Due to the decoupling argument, the global minimum is located in the $n_1=n_5=n_7=0$ subspace.
Within this subspace, the equipotential surfaces are cylinders along $n_3$, $n_8$ and are concentric spheres
in the $(n_2,n_4,n_6)$-subspace.

\begin{figure}[!htb]
   \centering
\includegraphics[width=6cm]{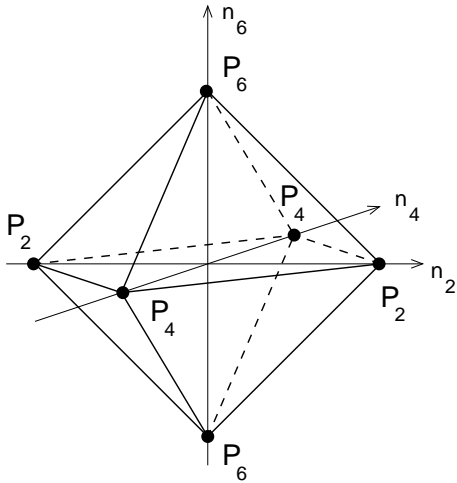}
\caption{Spontaneous breaking of the octahedral symmetry in the $(n_2, n_4, n_6)$-space. The octahedron drawn here
depicts the orbit space in this subspace, and it has slightly convex faces. The critical equipotential surface
not shown here is the large sphere that touches the six vertices of the octahedron.}
   \label{fig-octa}
\end{figure}

In the Appendix we show that the $n_1=n_5=n_7=0$ section of the orbit space, 
when projected into the $(n_2,n_4,n_6)$-space, has the shape of a slightly convex octahedron, 
which is schematically shown in Fig.~\ref{fig-octa}. The vertices and the edges are indeed those of
an octahedron, while the faces are not planar but slightly convex. 

Within the $(n_2,n_4,n_6)$-subspace, the octahedral symmetry group $O_h$ arises from permutation of the three axes $n_2, n_4, n_6$
as well as their sign flips. This group can be extended into the entire adjoint space. At the level
of Higgs doublets, it is generated by the following transformations:
the $CP$-transformation (which flips axes $n_2$, $n_5$, $n_7$), the sign flips of individual doublets, 
the $Z_2$ transformations such as $\phi_2 \leftrightarrow i \phi_3$,
and the $Z_3$ transformation $\phi_a \to R_{ab}\phi_b$ with matrix
\be
R_{ab} = 
\left(\begin{array}{ccc} 
0 & 1 & 0 \\ 
0 & 0 & -i \\ 
i & 0 & 0 \end{array}\right)\,,\quad 
\ee
which generates the cyclic changes $n_2 \to -n_6 \to n_4 \to n_2$ and $n_1 \to n_7 \to n_5 \to n_1$.
Clearly, the Higgs potential (\ref{octapotential}) is also symmetric under $O_h$.

The critical equipotential surface of the potential is the circumscribed sphere of the octahedron,
which touches it only at the six vertices labeled in Fig.~\ref{fig-octa} as $P_2$, $P_4$, $P_6$.
These vertices correspond to the neutral vacua with the following v.e.v.'s of the lower components
of the doublets:
\be
(\lr{\phi_1^0},\, \lr{\phi_2^0},\, \lr{\phi_3^0}) = {1\over\sqrt{2}}(v,\,\pm i v,\,0)\quad \mbox{or} 
\quad {1\over\sqrt{2}}(v,\,0,\, \pm v)
\quad \mbox{or}\quad {1\over\sqrt{2}}(0,\,v,\, \pm v)\,.\label{vacua-octahedral}
\ee
Thus, there are six isolated degenerate minima of the potential (\ref{octapotential}),
and at each of these minima the symmetry group $O_h$ is broken down to $D_4$.
An explicit expansion at any of these six minima gives the following physical Higgs boson masses:
\bea
h_i^\pm: && m^2_i = -{\lambda_{246} \over 2}v^2\,,\quad  -{\lambda_{246} \over 4}v^2\,,\\
h^0_i: && m^2_i = {\lambda_{157} \over 4}v^2\,,\quad {\lambda_{157} - \lambda_{246} \over 2}v^2\,,
\quad -{\lambda_{246} \over 2}v^2\,,\quad {4\lambda_0 + 3\lambda_{246} \over 6}v^2\,,
\quad {- \lambda_{246} \over 2}v^2\,.\nonumber
\eea
Notice that in (\ref{vacua-octahedral}) two minima out of six have relative phase between the two
non-zero vacuum expectation values. However this does not imply that these vacua are $CP$-violating
\footnote{I am thankful to Pedro Ferreira for this remark.}.
Indeed, these minima are physically equivalent to the other four, so that 
there exists a basis transformation from $O_h$ that rotates away the phase without changing altogether
the parameters of the potential. Thus, despite a relative phase, all six minima are $CP$-conserving.
Examples of this situation are also known in 2HDM, see e.g. \cite{lavoura1994}.
Whether 3HDM permits coexistence and/or degeneracy of $CP$-conserving and spontaneously 
$CP$-violating minima remains to be studied.

Similarly to the previous case, less symmetric situations can be easily devised by considering
the ellipsoidal rather than spherical equipotential surfaces.

\section{Discussion and conclusions}

The principal result of this paper is development of a way to efficiently 
treat the general NHDM Higgs potentials. We illustrated this approach 
with several particular 3HDM examples, which exhibited various symmetry breaking patterns.
We stress that these examples do not exhaust all the possibilities in 3HDM; 
however,  they revealed a number of interesting differences
between the scalar sectors of 2HDM and 3HDM.
In particular, we made the following observations about the minima and symmetries of 
3HDM potentials:
\begin{itemize}
\item
minima that spontaneously break and conserve a $Z_2$ symmetry can coexist and even be degenerate,
\item
neutral and charge-breaking minima can coexist (this was also noted in \cite{barroso2006}) and be degenerate,
\item 
there can be triple-degenerate charge-breaking vacua and six-fold degenerate neutral minima,
\item
degenerate minima do not necessarily imply that the potential itself has any symmetry. 
\end{itemize}
The last statement comes from the simple observation that we can break the $Z_3$-symmetry of the potential 
(\ref{Z3explicit}) by drawing an ellipse in the root plane and placing it asymmetrically 
with respect to the triangle, so that it still passes through all three vertices of the triangle. 
The three neutral points will then be the degenerate global minima of the potential, although with distinct
Higgs mass boson masses. This should be contrasted to 2HDM, where in the tree-level approximation 
degenerate minima can occur only as a result of spontaneous breaking of some symmetry, \cite{ivanov1}.

In principle, all the specific 3HDM examples we considered could be easily analyzed with straightforward
algebra, which we indeed used to calculate the Higgs boson masses. 
However, it is through the geometric methods that these examples were found.
The same geometric arguments also guaranteed that the minima we considered were global.
However, the power of geometric approach goes beyond such simple examples, as one can
easily imagine less symmetric types of contact between the critical equipotential surface 
and the orbit space.

For example, in the symmetric cases we had several degenerate global minima in the orbit space. 
However this degeneracy can always be lifted by a sufficiently small non-symmetric shift
of the critical equipotential surface. In this way we can get, for example, 
a potential with the charge-breaking minimum lying slightly higher or lower 
than the neutral one, or a potential with several neutral minima lying at different depths.

Let us also note that in all the cases we considered the degenerate minima were located 
at different $r_i$ but equal $r_0$. This property can be easily broken
if a sufficiently skewed (due to non-diagonal $\Lambda_{\mu\nu}$) 
or displaced (due to non-zero $M_i$) equipotential surfaces are considered.
However, the experience from 2HDM lends support to the argument that such skewed or displaced
cases will not generate any new pattern of symmetry breaking in addition to what can be generated with more 
symmetrically placed equipotential surfaces.

Let us also comment on the number of degenerate minima in 3HDM.
Due to the Bezout theorem in algebraic geometry, a second-order (the critical equipotential surface) 
and a third-order (the charge-breaking orbit space) surfaces intersect along an algebraic manifold 
with order less or equal to six. If this intersection is meant to be a union 
of disjoint points, one needs to allow two orders per point, giving thus three points.
Therefore, the charge-breaking global minimum has degeneracy in 3HDM not more than three.
Our example with spontaneously broken $Z_3$ realizes this maximally degenerate symmetry breaking
to the charge-breaking vacuum.

The neutral orbit space is represented by an intersection of the same cubic surface and
the forward lightcone, which is a quadric. Therefore, the neutral orbit itself is a sextic manifold.
Its intersection with the orbit space has order 12, which can encode up to six
separate points in the orbit space. Therefore, the neutral order space can have degeneracy up to six.
The six-fold degenerate neutral vacuum was realized in the example with the octahedral symmetry. 

One can think of several directions for future work.

Perhaps, the first goal would be to give a complete classification of explicit symmetries 
of the NHDM potential for different $N$.
Especially interesting are the discrete symmetries, whose spontaneous breaking does not lead to
extra massless scalars. One can study phenomenological manifestations of 
various discrete symmetries and check to what extent these symmetries can be incorporated 
into the Yukawa sector and what symmetries in the fermion masses and mixing can result.

Then, one should also try refining the algebraic description of the Higgs sector in NHDM.
This includes a further study of properties of the NHDM orbit space, both its local structure and topology
beyond what was analyzed in \cite{NHDM2010},
a derivation of compact algebraic criteria for the positivity constraints on the Higgs potential,
a derivation of basis-invariant criteria for existence of a hidden symmetry, etc.
It would be also very interesting to see whether the Minkowski-space formalism of \cite{ivanov1,ivanov2,nishi2008}
can be as efficient in NHDM as it was in 2HDM.

In the case of general 2HDM, some dynamical properties such as the mass matrix of the physical Higgs bosons
\cite{oneil,IvanovDegee} and thermal evolution of the vacuum Higgs configuration \cite{thermal} have also been addressed.
It is interesting to see if these calculations can be repeated for the general $N$.

Finally, we hope that the general approach of this paper, 
based on the equipotential surface technique in the orbit space, can be generalized to other Higgs models,
involving singlets, triplets etc.
\\

In summary, continuing our study of the $N$-Higgs-doublet model initiated in \cite{NHDM2010},
we developed a method which helped us analyze the properties of NHDM Higgs potentials. 
We worked in the space of Higgs field bilinears,
and the key roles in our approach were played by certain geometric constructions in this space, 
such as the equipotential surfaces and the orbit space.
We also studied explicit symmetries possible in NHDM and their spontaneous breaking.
We illustrated the general discussion with various particular cases of 3HDM,
which revealed a wealth of possibilities which were not realizable in 2HDM.
The technique suggested here can be used to gain more further insight into 
properties of multi-Higgs-doublet models.

\section*{Acknowledgements} 
This work has grown from very stimulating discussions 
with P.~Ferreira, C.~Nishi and J.~Silva. I am also thankful to the remarks they made 
after reading this manuscript.
The support of the Belgian Fund F.R.S.-FNRS via the
contract of Charg\'e de recherches, and it part of grants
RFBR No.08-02-00334-a and NSh-3810.2010.2 is also acknowledged.

\appendix

\section{Orbit space in the transverse space: specific 3HDM examples}

In Section~\ref{section:3HDM-examples} we considered several examples of symmetries in 3HDM
coming from the ``transverse space'', that is from the space orthogonal to the root plane.
Here we derive the orbit space shapes in two specific cases:
taking the $n_1=n_2=n_5=n_7=0$ section of the orbit space and projecting
it onto the $(n_4,n_6)$ plane, and taking the $n_1=n_5=n_7=0$ section and 
projecting it into the $(n_2,n_4,n_6)$ space.
We use these projections (disregarding the values of $n_3$, $n_8$) 
because the $D_4$- and $O_h$-symmetric potentials 
considered in Section~\ref{section:3HDM-examples} 
do not depend on $n_3$ and $n_8$.

We start with the $d$-condition (\ref{d:bari}) and the lightcone condition
\be
\label{LC:bari}
\vec n^2 = |n_{12}|^2+ |n'_{12}|^2 + |n''_{12}|^2 + { 3(p^2 + p^{\prime 2} + p^{\prime\prime 2}) - 1\over 2} \le 1\,.
\ee
From the fact the any $2\times 2$ minor of the $K$-matrix is non-negative, we derive that
\be
\label{zab:bari}
3 p'p'' \ge |n_{12}|^2\,,\quad 3 p p' \ge |n''_{12}|^2\,,\quad  3 p p'' \ge |n'_{12}|^2\,.
\ee 
Let us now restrict ourselves to the subspace $n_1=n_2=n_5=n_7=0$. We are interested in the projection
of the orbit space onto $(n_4,n_6)$. The $d$-condition then simplifies to
\be
\label{d:n12-2}
p' n_6^2 + p'' n_4^2 = 3 p p' p''\,.
\ee
Let us first consider the cases when at least one among $p$, $p'$, and $p''$ is zero.
If $p=0$, then $|n''_{12}|^2 = |n'_{12}|^2 = 0$, and we return to the root plane staying on the upper edge of the triangle.
After projecting from the $(n_3,n_4,n_6,n_8)$ space to the $(n_4,n_6)$ plane, it corresponds to the point at the origin.

If instead $p'=0$, then $n''_{12} = 0$, so in the ``transverse'' space the only non-zero component is $n_6$,
whose value is 
\be
n_6^2 \le  3 p p'' = 3 p(1-p)\,.
\ee
As $p$ changes from zero to one, 
the value of $n_6^2$ is limited to $3/4$, and this particular point corresponds to the neutral vacuum. 
Thus, on the $(n_4,n_6)$-plane we get
a line segment along the $n_4=0$ axis from $-\sqrt{3}/2$ to $\sqrt{3}/2$.
A similar segment along the $n_6=0$ axis arises in the case $p''=0$.

Let us now consider the case when all $p$, $p'$, and $p''$ are non-zero (i.e. the projection on the root plane gives
a point strictly inside the triangle). Due to $3 p' p'' > |n_{12}|^2 = 0$, we necessarily stay on the charge-breaking manifold.
Then (\ref{d:n12-2}) can be rewritten as
\be
{n_6^2 \over 3 p p''} + {n_4^2 \over 3 p p'} = 1\,.
\ee
Thus, for each point inside the triangle defined by $p$, $p'$, and $p''$, we get an ellipse in the $(n_4,n_6)$ plane
with semiaxes 
\be
a_4^2 = 3 p p'\,,\ a_6^2 = 3 p p''\,. 
\ee
Note that $a_4^2 + a_6^2 = 3 p(1-p)$; thus the largest possible ellipse with a given eccentricity
is at $p=1/2$. The semiaxes of this largest ellipse can be parametrized as 
$$
a_4 = {\sqrt{3} \over 2}\cos\theta\,,\quad a_6 = {\sqrt{3} \over 2}\sin\theta\,,\quad 0\le \theta \le {\pi\over 2}\,. 
$$
We then get a one-parametric family of the largest ellipses with different values of $\theta$.
If the angular coordinate is $\alpha$ in each ellipse, then
\be
n_4 = {\sqrt{3} \over 2}\cos\theta\cos\alpha\,,\quad 
n_6 = {\sqrt{3} \over 2}\sin\theta\sin\alpha\,.
\ee
Thus, the family of ellipses sweep the square on the $(n_4,n_6)$-plane, bounded by
\be
n_4 \pm n_6 = {\sqrt{3} \over 2}\cos(\theta \mp\alpha) \in \left[-{\sqrt{3} \over 2},{\sqrt{3} \over 2}\right]\,.
\ee
Thus, we recover the square of Fig.~\ref{fig-d4}:
the vertices and the edges of the triangle correspond to the neutral and charge-breaking points, respectively,
while the interior can corresponds to the charge-breaking orbit space or, if a point lies on an axis, 
to the neutral orbit space.

Consider now another section of the orbit space $n_1=n_5=n_7=0$; we are now interested in the projection
of the orbit space in the $(n_2,n_4,n_6)$-space.
We selected these axes in such a way that $\Re(n_{12}n'_{12}n''_{12}) = \Re(i n_2 n_4 n_6) = 0$.
Thus, the $d$-condition simplifies to
\be
\label{d:157-1}
p n_2^2 + p' n_6^2 + p'' n_4^2 = 3 p p' p''\,.
\ee
Again, a similar analysis shows that the neutral orbit space corresponds to a ``cross'', that is,
three line segments spanning from $-\sqrt{3}/2$ to $\sqrt{3}/2$ along each of the three coordinate axes.
If all $p$'s are non-zero, we can rewrite the $d$-condition as
\be
{n_2^2 \over 3 p' p''} + {n_6^2 \over 3 p p''} + {n_4^2 \over 3 p p'} = 1\,,
\ee
which, at given $p$'s, parametrizes an ellipsoid in the $(n_2,n_4,n_6)$-space.
Considering all the $p$'s inside the triangle gives us a two-parametric family of ellipsoids
which sweep a certain region $(n_2,n_4,n_6)$-space. This region has square sections
in the planes $n_2=0$, $n_4=0$, or $n_6=0$, and it has the symmetry of an octahedron.
It is convex but the curvature of the faces is sufficiently small, so that
this region lies inside a sphere with radius $\sqrt{3}/2$, touching it only at the six vertices
corresponding to the neutral orbit space, see Fig.~\ref{fig-octa}.

\end{document}